%% file: alife12.tex
\documentclass[letterpaper]{article}
\usepackage{natbib,alifexii,pstricks}
\title{Network Complexity of Foodwebs}
\author{Russell Standish\\School of Mathematics and Statistics,
  University of New South Wales}
\newcommand{\EcoLab}{{\sffamily\slshape
    \mbox{\raisebox{.5ex}{Eco}\hspace{-.4em}{\makebox[.5em]{L}ab}}}}

\begin{document}
\maketitle

\begin{abstract}
In previous work, I have developed an information theoretic complexity
measure of networks. When applied to several real world food webs,
there is a distinct difference in complexity between the real food
web, and randomised control networks obtained by shuffling the network links.
One hypothesis is that this complexity surplus represents
information captured by the evolutionary process that generated the
network. 

In this paper, I test this idea by applying the same complexity
measure to several well-known artificial life models that exhibit
ecological networks: Tierra, EcoLab and Webworld. Contrary to what was
found in real networks, the artificial life generated foodwebs had
little information difference between itself and randomly shuffled versions.

\end{abstract}

\section{Introduction}

In \cite{Standish05a}, I developed a method for computing the
information complexity of a network. In \cite{Standish10a}, I refined
and generalised the method to overcome a problem with higher
complexity values of empty and full networks relative to partially
filled networks of the same degree, as well as taking account of link
weights. Coupled with some new algorithms for computing automorphism
group size, this network complexity measure is practical for networks
of several thousand nodes.

In \cite{Standish10a}, I studied several published datasets of natural
networks, including a number of foodwebs available from the Pajek
website, and the neural network of {\em C. elegans} (see Table
\ref{complexity data}). In most cases, these networks exhibited
significantly heightened complexity values compared with those of
control networks obtained by shuffling the links in a random fashion.
This leads to the hypothesis that evolutionary processes tend to
produce networks with a {\em complexity surplus} ($\Delta$) compared
with random assembly processes.

\begin{table*}
\begin{center}
\begin{tabular}{lrrrrrr}
Dataset & nodes & links &${\cal C}$ & $e^{\langle\ln{\cal
  C}_\mathrm{ER}\rangle}$ & $\Delta = {\cal
  C}-e^{\langle\ln{\cal C}_\mathrm{ER}\rangle}$ & $\frac{|\ln{\cal
  C}-\langle\ln{\cal C}_\mathrm{ER}\rangle|}{\sigma_\mathrm{ER}}$ \\\hline
celegansneural & 297 & 2345 & 442.7 &251.6 &191.1 &29\\
celegansmetabolic & 453 & 4050 & 25421.8 & 25387.2 & 34.6&$\infty$  \\
lesmis & 77 & 508 &199.7 &114.2 &85.4 &24\\
adjnoun & 112 & 850 & 3891 & 3890 & 0.98 & $\infty$\\
yeast & 2112 & 4406 & 33500.6 & 30218.2 & 3282.4 & 113.0\\
baydry & 128 & 2138 &126.6 &54.2 &72.3 &22\\
baywet & 128 & 2107 &128.3 &51.0 &77.3 &20\\
cypdry & 71 & 641 & 85.7 &44.1 &41.5 &13\\
cypwet & 71 & 632 & 87.4 &42.3 &45.0 &14\\
gramdry & 69 & 911 & 47.4 &31.6 &15.8 &10\\
gramwet & 69 & 912 &54.5 &32.7 &21.8 &12\\
Chesapeake& 39& 177& 66.8& 45.7& 21.1& 10.4\\
ChesLower& 37& 178& 82.1& 62.5& 19.6& 10.6\\
ChesMiddle& 37& 208& 65.2& 48.0& 17.3& 9.3\\
ChesUpper& 37& 215& 81.8& 60.7& 21.1& 10.2\\
CrystalC& 24& 126& 31.1& 24.2& 6.9& 6.4\\
CrystalD& 24& 100& 31.3& 24.2& 7.0& 6.2\\
Everglades &69& 912& 54.5& 32.7& 21.8& 11.8\\
Florida& 128& 2107& 128.4& 51.0& 77.3& 20.1\\
Maspalomas& 24& 83& 70.3& 61.7& 8.6& 5.3\\
Michigan& 39& 219& 47.6& 33.7& 14.0& 9.5\\
Mondego& 46& 393& 45.2& 32.2& 13.0& 10.0\\
Narragan& 35& 219& 58.2& 39.6& 18.6& 11.0\\
Rhode& 19& 54& 36.3& 30.3& 6.0& 5.3\\
StMarks& 54& 354& 110.8& 73.6& 37.2& 16.0\\
\end{tabular}
\end{center}
\caption{
  Complexity values of several freely available network
  datasets, as reported in \cite{Standish10a}.  For
  each network, the number of 
  nodes and links are given, along with the computed complexity ${\cal
  C}$. In the fourth column, the original network is shuffled 1000
  times, and the logarithm of the complexity is averaged
  ($\langle\ln{\cal C}_\mathrm{ER}\rangle$). The fifth
  column gives the difference between these two values, which
  represents the information content of the specific arrangement of
  links. The final column gives a measure of the significance of this
  difference in terms of the number of standard deviations (``sigmas'') of the
  distribution of shuffled networks. In two examples, the
  distribution of shuffled networks had zero standard deviation, so
  $\infty$ appears in this column.
}
\label{complexity data}
\end{table*}

In this work, I apply the same methods to networks created by
artificial life evolutionary systems, in particular the interaction
network of Tierra \citep{Ray91} and the foodwebs of
EcoLab \citep{Standish94} and Webworld \citep{Caldarelli-etal98}.

\section{Complexity as Information}

The notion of using information content as a complexity measure is
fairly simple.  In most cases, there is an obvious {\em prefix-free}
representation language within which descriptions of the objects of
interest can be encoded.  There is also a classifier of descriptions
that can determine if two descriptions correspond to the same object.
This classifier is commonly called the {\em observer}, denoted $O(x)$.

To compute the complexity of some object $x$, count the number of
equivalent descriptions $\omega(\ell,x)$ of length $\ell$ that map to
the object $x$ under the agreed classifier. Then the complexity of $x$
is given in the limit as $\ell\rightarrow\infty$:
\begin{equation}\label{complexity}
{\cal C}(x) = \lim_{\ell\rightarrow\infty} \ell\log N - \log\omega(\ell,x)
\end{equation}
where $N$ is the size of the alphabet used for the representation language.

Because the representation language is prefix-free, every description
$y$ in that language has a unique prefix of length $s(y)$.  The
classifier does not care what symbols appear after this unique prefix.
Hence $\omega(\ell,O(y))\geq N^{\ell-s(y)}$. As $\ell$ increases,
$\omega$ must increase as fast, if not faster than $N^\ell$, and do so
monotonically. Therefore $C(O(y))$ decreases monotonically with
$\ell$, but is bounded below by 0. So equation (\ref{complexity})
converges.

To use this formalism with networks, we need to fix two things: how to
decide when two networks are identical, and a prefix-free
representation language, which will be used to count the
representations of a given network. In this context, ignoring any link
weights, two networks are considered identical if the nodes of one can
be placed over the nodes of the second one, such that the links
correspond exactly. They are topologically identical. We ignore any
labels on the nodes or links.

\subsection{Network bitstring representation}

To represent the network as a bitstring, we need to store the node
count ($n$) and link count ($l$), as well as representation of the adjacency
matrix. The initial part of the string has $w=\lceil \log_2 n\rceil$ `1'
bits, followed by a single `0' stop bit. Following that are $w$ bits
representing the value of $n$ in binary. Knowing the value of $n$,
the number of bits needed to represent $l$ is $\lceil \log_2 L
\rceil$, where $L=(n(n-1)/2)$ so $l$ is stored in a field of that width. 

For the final part of the string, the linkfield, we can represent the
adjacency matrix such that a `1' bit in position $i(n-1)+j$-th
represents a link from node $i$ to $j$ if $j<i$ or from $i$ to $j+1$
if $j>i$, where nodes are numbered $0\ldots n-1$, $i<n$ and $j<n-1$.
However, this representation is not efficient --- given $l$, there
must be exactly $l$ `1' bits in the linkfield, ie it is one of the
permutations of $l$ `1' bits and $L-l$ `0' bits. We can
enumerate the 
$\left(\begin{array}{c} L \\ l \end{array}\right)$
permutations, and choose the rank of our linkfield in the enumeration
as the encoding of the linkfield. This is known as rank encoding
\citep{Myrvold-Ruskey01}. One of the effects of choosing this encoding
is that both an empty and a full network have just one possible
linkfield, so will have a rank encoding of 0, representable
in 0 bits, as we already know whether a network is empty or full from
the values of $n$ and $l$. Hence, the full and empty networks are the
simplest networks for given $n$ and $l$.

\subsection{Weighted links}

Whilst the information contained in link weights might be significant
in some circumstances (for instance the weights of a neural network
can only be varied in a limited range without changing the overall
qualitative behaviour of the network), of particular theoretical
interest is to consider the weights as continuous parameters
connecting one network structure with another. For instance if a
network $X$ has the same network structure as A, with $b$ links of
weight 1 with a network structure $B$ and the remaining $a-b$ links of
weight $w$, then we would like the network complexity of $X$ to vary
smoothly between that of $A$ and $B$ as $w$ varies from 1 to
0. \cite{Gornerup-Crutchfield08} introduced a similar measure.

The most obvious way of defining this continuous complexity measure is
to start with normalised weights $\sum_iw_i=1$. Then arrange the links
in weight order, and compute the complexity of networks with just
those links of weights less than $w$. The final complexity value of a
network $X = N\times L$, where $N$ is the set of nodes, and $L$ the
set of links with associated weights $w_i, \exists i\in L$, is
obtained by integrating:
\begin{equation}\label{weighted C}
{\cal C}(X=N\times L) = \int_0^1 {\cal C}(N\times\{i\in L: w_i<w\}) dw
\end{equation}
Obviously, since the integrand is a stepped function, this is computed
in practice by a sum of complexities of partial networks.

\subsection{Counting the representations}

In principle, one could compute the complexity of a network by
enumerating all bitstrings for a given $n$ and $l$, and counting the
number of bitstrings that represent the target network. However, this
algorithm is highly combinatoric, and only really feasible for small
networks. However, the number of representations can also be computed
by dividing the total number of possible renumberings of the nodes
($N!$) by the size of the automorphism group, for which several
practical algorithms
exist \citep{McKay81,Standish09a,Darga-etal08}. Even though each of
these algorithms is NP-complete, in practice they tend to perform
quite well for networks up to several thousands of nodes. Where each
algorithm performs poorly, one of the other algorithms performs
well, so a hybrid algorithm that runs each algorithm in parallel, and
returning the result of the first algorithm to complete, performs
extremely well. 

\section{ALife models}

\subsection{Tierra}\label{tierra}

Tierra \citep{Ray91} is a well known artificial life system in which
self reproducing computer programs written in an assembly-like
language are allowed to evolve. The programs, or {\em digital
  organisms} can interact with each via template matching operations,
modelled loosely on the way proteins interact in real biological
systems. A number of distinct strategies evolve, including parasitism,
where organisms make use of another organism's code and
hyper-parasitism where an organism sets traps for parasites in order
to steal their CPU resources. At any point in time in a Tierra run,
there is an interaction network between the species present, which is
the closest thing in the Tierra world to a foodweb.

Tierra is an aging platform, with the last release (v6.02) having been
released more than six years ago. For this work, I used an even older
release (5.0), for which I have had some experience in working
with. Tierra was originally written in C for an environment where ints
were 16 bits and long ints 32 bits. This posed a problem for using it
on the current generation of 64 bit computers, where the word sizes
are doubled. Some effort was needed to get the code 64 bit
clean. Secondly a means of extracting the interaction network was
needed. Whilst Tierra provided the concept of ``watch bits'', which
recorded whether a digital organism had accessed another's genome or
vice versa, it did not record which other genome was accessed. So I
modified the template matching code to log the pair of genome
labels that performed the template match to a file.

Having a record of interactions by genotype label, it is necessary to map the
genotype to phenotype. In Tierra, the phenotype is the behaviour of
the digital organism, and can be judged by running the organisms
pairwise in a tournament, to see what effect each has on the
other. The precise details for how this can be done is described in
\cite{Standish03a}.  

Having a record of interactions between phenotypes, and discarding
self-self interactions, there are a number of ways of turning that
record into a foodweb. The simplest way, which I adopted, was sum the
interactions between each pair of phenotypes over a sliding window of
100 million executed instructions, and doing this every 20 million executed
instructions. This lead to time series of around 2000 foodwebs for
each Tierra run. 

In Tierra, parsimony pressure is controlled by the parameter
SlicePow. CPU time is allocated proportional to genome size raised to
SlicePow. If SlicePow is close to 0, then there is great evolutionary
pressure for the organisms to get as small as possible to increase
their replication rate. When it is one, this pressure is
eliminated. In \cite{Standish04c}, I found that a SlicePow of around
0.95 was optimal. If it were much higher, the organisms grow so large and so
rapidly that they eventually occupy more than 50\% of the soup. At
which point they kill the soup at their next Mal (memory allocation)
operation. In this work, I altered the implementation of Mal to fail
if the request was more than than the soup size divided by minimum
population save threshold (usually around 10). Organisms any larger
than this will never appear in the Genebanker (Tierra's equivalent of
the fossil record), as their population can never exceed the save
threshold. This modification allows SlicePow = 1 runs to run for
an extensive period of time without the soup dying. 

\subsection{EcoLab}

EcoLab was introduced by the author as a simple model of an evolving
ecosystem \citep{Standish94}. The ecological dynamics is described by an
$n$-dimensional generalised Lotka-Volterra equation:
\begin{equation}
\dot{n_i} = r_in_i + \sum_j\beta_{ij}n_in_j,
\end{equation}
where $n_i$ is the population density of species $i$, $r_i$ its growth
rate and $\beta_{ij}$ the interaction matrix. Extinction is handled
via a novel stochastic truncation algorithm, rather than the more usual
threshold method. Speciation occurs by randomly mutating th ecological
parameters ($r_i$ and $\beta_{ij}$) of the parents, subject to the
constraint that the system remain bounded \citep{Standish98b}.

The interaction matrix is a candidate foodweb, but has too much
information. Its offdiagonal terms may be negative as well as
positive, whereas for the complexity definition (\ref{weighted C}), we
need the link weights to be positive. There are a number of ways of
resolving this issue, such as ignoring the sign of the off-diagonal
term (ie taking its absolute value), and antisymmetrising the matrix by
subtracting its transpose, then using the sign of the offdiagonal term
to determine the link direction.

For the purposes of this study, I chose to subtract just the negative
$\beta_{ij}$ terms from itself and its transpose term $\beta_{ji}$.
This effects a maximal encoding of the interaction matrix information in
the network structure, with link direction and weight encoding the
direction and size of resource flow. The effect is as follows:
\begin{itemize}
\item Both $\beta_{ij}$ and $\beta_{ji}$ are positive (the {\em
    mutualist} case). Neither offdiagonal term changes, and the two nodes
    have links pointing in both directions, with weights given by the
    two offdiagonal terms.
\item Both $\beta_{ij}$ and $\beta_{ji}$ are negative (the {\em
    competitive} case). The terms are swapped, and the signs changed
    to be positive. Again the two nodes have links pointing in both
    directions, but the link direction reflects the direction of
    resource flow.
\item Both $\beta_{ij}$ and $\beta_{ji}$ are of opposite sign (the
  {\em predator-prey} or {\em parasitic} case). Only a single link
  exists between species $i$ and $j$, whose weight is the summed
  absolute values of the offdiagonal terms, and whose link direction
  reflects the direction of resource flow.
\end{itemize}

\subsection{Webworld}

Webworld is another evolving ecology model, similar in some respects
to EcoLab, introduced by \cite{Caldarelli-etal98}, with some
modifications described in \cite{Drossel-etal01}. It features more
realistic ecological interactions than does EcoLab, in that it tracks
biomass resources. It too has an interaction matrix called a {\em
  functional response} in that model that could serve as a foodweb,
which is converted to a directed weighted graph in the same way as the
EcoLab interaction matrix. I used the Webworld implementation
distributed with the \EcoLab{} simulation platform \cite{Standish04a}.

\section{Results}

\begin{figure*}
\small
\begin{center}
\input{SlicePow0.95}\\
Instructions Executed ($\times10^{10}$)
\end{center}
\caption{Complexity of the Tierran interaction network for
  SlicePow=0.95, and $\Delta$, exaggerated by a factor of 100. Two
  different random number generators were used, Havege and the normal
  linear congruential generator supplied with Tierra. }
\label{start-fig}
\end{figure*}
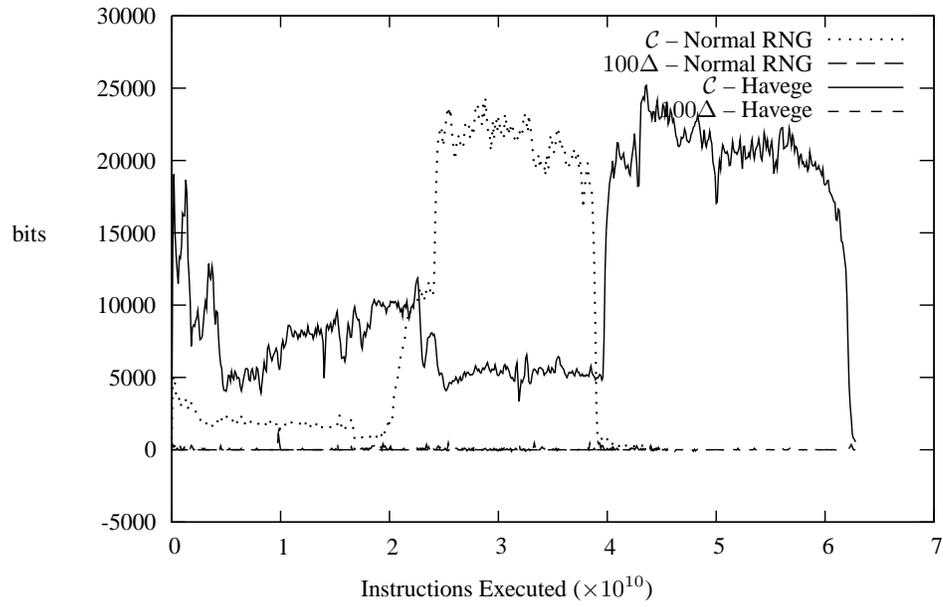

\begin{figure*}
\small
\begin{center}
\input{SlicePow1}\\
Instructions Executed ($\times10^{10}$)
\end{center}
\caption{Complexity of the Tierran interaction network for
  SlicePow=1, and $\Delta$, exaggerated by a factor of 100. Two
  different random number generators were used, Havege and the normal
  linear congruential generator supplied with Tierra.}
\end{figure*}

\begin{figure*}
\small
\begin{center}
\input{ecolab.complexity}\\
Timesteps ($\times10^{7}$)
\end{center}
\caption{Complexity of EcoLab's foodweb, and $\Delta$, exaggerated by a factor of 100, as described in the text.  }
\end{figure*}

\begin{figure*}
  \small
\begin{center}
\input{webworld.complexity}\\
Timesteps ($\times10^{7}$)
\end{center}
\caption{Complexity of Webworld's foodweb, and $\Delta$, exaggerated by a factor of 100, as described in the text.  }
\label{end-fig}
\end{figure*}

\section{Methods and materials}

Tierra was run on a 512KB soup, with SlicePow set to 1, until the soup
died, typically after some $5\times10^{10}$ instructions have executed. Some
variant runs were performed with SlicePow=0.95, and with different random
number generators, but no difference in the outcome was observed.

The source code of Tierra 5.0 was modified in a few places, as
described in the Tierra section of this paper. The final source code
is available as tierra.5.0.D7.tar.gz from the \EcoLab{} website hosted
on SourceForge (http://ecolab.sf.net).

The genebanker output was processed by the eco-tierra.3.D13 code, also
available from the \EcoLab{} website, to produce a list of phenotype
equivalents for each genotype. A function for processing the
interaction log file generated by Tierra and producing a timeseries of
foodweb graphs was added to Eco-tierra. The script for running this
postprocessing step is process\_ecollog.tcl.

The EcoLab model was adapted to convert the interaction matrix into a
foodweb and log the foodweb to disk every 1000 time steps for later
processing. The Webworld model was adapted similarly. The model parameters
were as documented in the included ecolab.tcl and webworld.tcl experiment files
of the ecolab.4.D37 distribution, which is also available from the \EcoLab{}
website.

Finally, each foodweb, and 100 link-shuffled control versions were run
through the network complexity algorithm (\ref{weighted C}). This is
documented in the cmpERmodel.tcl script of ecolab.4.D37.  The average
and standard deviation of $\ln {\cal C}$ was calculated, rather than
${\cal C}$ directly, as the shuffled complexity values fitted a
log-normal distribution better than a standard normal distribution.
The difference between the measured complexity and $\exp\langle\ln
{\cal C}\rangle$ (ie the geometric mean of the control network
complexities) is what is reported as $\Delta$ in Figures
\ref{start-fig}--\ref{end-fig}.

\section{Discussion}

It can be seen from Figures \ref{start-fig}--\ref{end-fig}, that none
of the artificial life models studied generate substantially greater
network complexities than do the control networks. By
``substantially'', I mean more than 10\% of the total network
complexity. The complexity difference that exists is nevertheless
often statistically significant, albeit small (of the order of a few
bits). By contrast, most of the 26 practical networks studied in
\cite{Standish10a} exhibited substantially greater complexities than
their controls, the exceptions being the David Copperfield adjective-noun
adjacency dataset (0.98 bits), and the {\em C. elegans} metabolic
network (which at 34.6 bits is about 0.1\% of the total complexity).

The complete failure for several independent artificial evolutionary
systems to be able to generate this complexity surplus weakens the
case for the surplus as being due to operation of an evolutionary
process. It is possible that this is another illustration of the
difference between artificial evolutionary systems and natural
evolutionary systems observed with Bedau-Packard statistics
\citep{Bedau-etal98}. There is also the possibility that some
systematic artifact skews the observational data towards more
symmetric networks (which increases complexity values), however it
seems implausible that networks collected by many different observers
in many different fields should exhibit the same systematic error.
More work needs to be done applying this complexity metric to both
artificially evolved networks and observational data of naturally
evolved networks to elucidate if this is artifact, or a real
phenomenon.

\section{Conclusion}

In this work, I measured the network complexity of several
artificially evolved foodwebs to see if I could reproduce the
complexity surplus seen in empirical network data. In none of the
artificial systems I studied was the complexity surplus substantial
enough to be considered a real effect.

\footnotesize
\bibliographystyle{apalike}
\bibliography{rus}

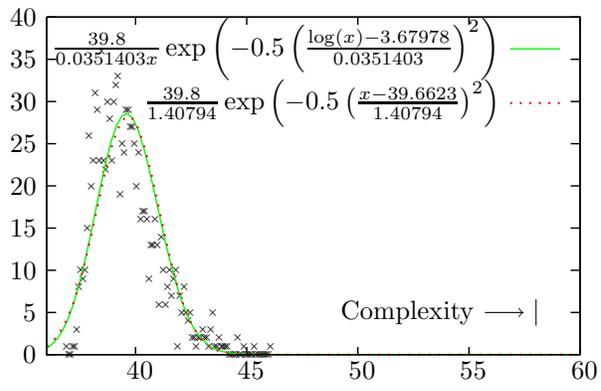
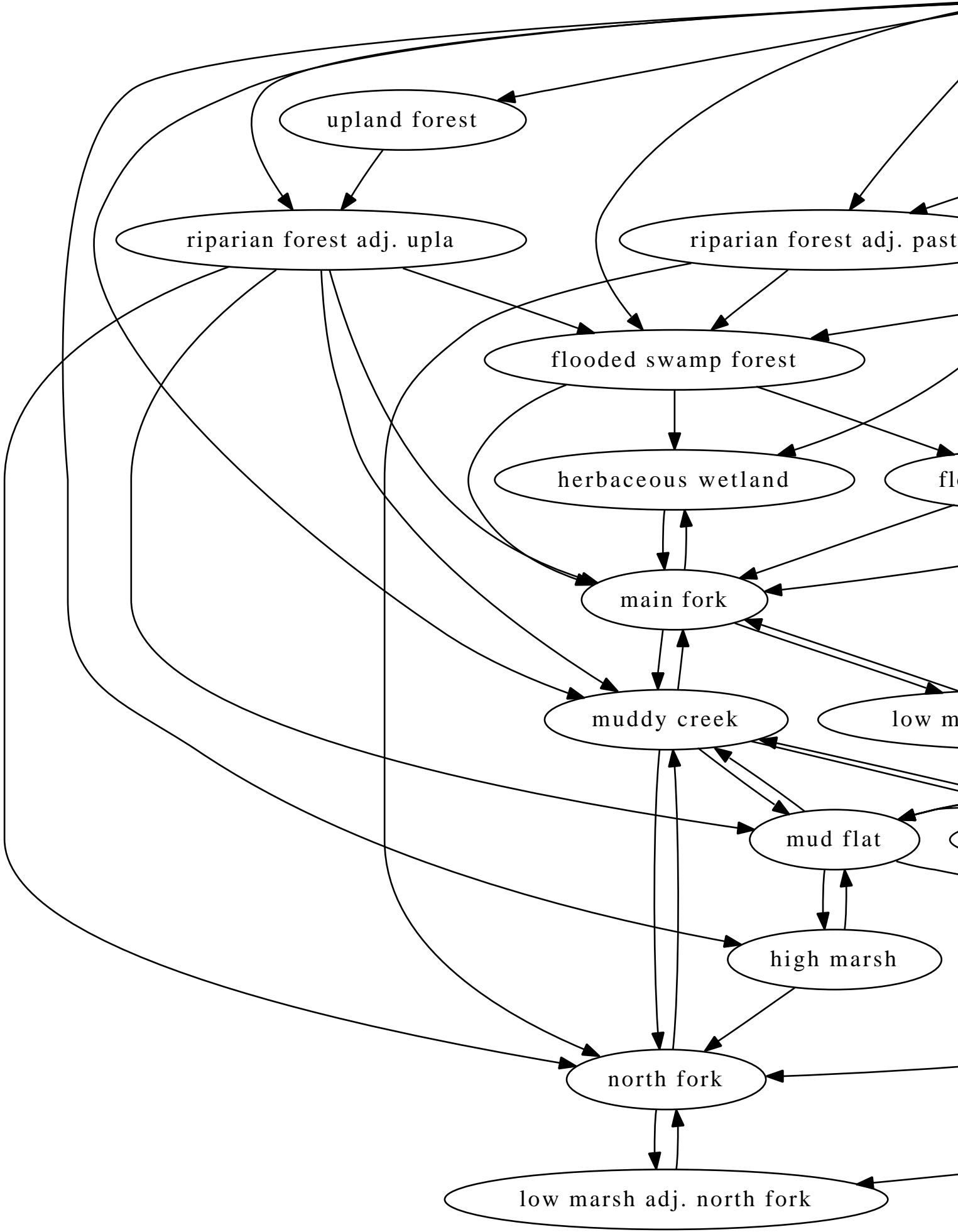
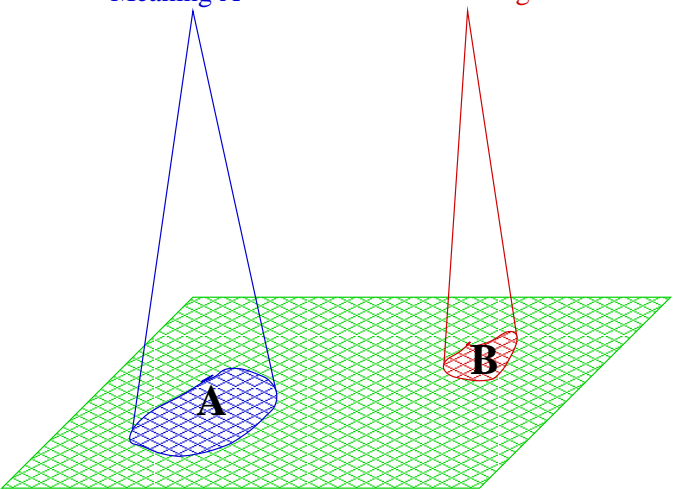
\end{document}

%% file: SlicePow0.95.tex
\ifx\PSTloaded\undefined
\def\PSTloaded{t}
\psset{arrowsize=.01 3.2 1.4 .3}
\psset{dotsize=.01}
\catcode`@=11

\newpsobject{PST@Border}{psline}{linewidth=.0015,linestyle=solid}
\newpsobject{PST@Axes}{psline}{linewidth=.0015,linestyle=dotted,dotsep=.004}
\newpsobject{PST@Solid}{psline}{linewidth=.0015,linestyle=solid}
\newpsobject{PST@Dashed}{psline}{linewidth=.0015,linestyle=dashed,dash=.01 .01}
\newpsobject{PST@Dotted}{psline}{linewidth=.0025,linestyle=dotted,dotsep=.008}
\newpsobject{PST@LongDash}{psline}{linewidth=.0015,linestyle=dashed,dash=.02 .01}
\newpsobject{PST@Diamond}{psdots}{linewidth=.001,linestyle=solid,dotstyle=square,dotangle=45}
\newpsobject{PST@Filldiamond}{psdots}{linewidth=.001,linestyle=solid,dotstyle=square*,dotangle=45}
\newpsobject{PST@Cross}{psdots}{linewidth=.001,linestyle=solid,dotstyle=+,dotangle=45}
\newpsobject{PST@Plus}{psdots}{linewidth=.001,linestyle=solid,dotstyle=+}
\newpsobject{PST@Square}{psdots}{linewidth=.001,linestyle=solid,dotstyle=square}
\newpsobject{PST@Circle}{psdots}{linewidth=.001,linestyle=solid,dotstyle=o}
\newpsobject{PST@Triangle}{psdots}{linewidth=.001,linestyle=solid,dotstyle=triangle}
\newpsobject{PST@Pentagon}{psdots}{linewidth=.001,linestyle=solid,dotstyle=pentagon}
\newpsobject{PST@Fillsquare}{psdots}{linewidth=.001,linestyle=solid,dotstyle=square*}
\newpsobject{PST@Fillcircle}{psdots}{linewidth=.001,linestyle=solid,dotstyle=*}
\newpsobject{PST@Filltriangle}{psdots}{linewidth=.001,linestyle=solid,dotstyle=triangle*}
\newpsobject{PST@Fillpentagon}{psdots}{linewidth=.001,linestyle=solid,dotstyle=pentagon*}
\newpsobject{PST@Arrow}{psline}{linewidth=.001,linestyle=solid}
\catcode`@=12

\fi
\psset{unit=5.0in,xunit=5.0in,yunit=3.0in}
\pspicture(0.000000,0.000000)(1.000000,1.000000)
\ifx\nofigs\undefined
\catcode`@=11

\PST@Border(0.1490,0.0840)
(0.1640,0.0840)

\PST@Border(0.9470,0.0840)
(0.9320,0.0840)

\rput[r](0.1330,0.0840){-5000}
\PST@Border(0.1490,0.2103)
(0.1640,0.2103)

\PST@Border(0.9470,0.2103)
(0.9320,0.2103)

\rput[r](0.1330,0.2103){ 0}
\PST@Border(0.1490,0.3366)
(0.1640,0.3366)

\PST@Border(0.9470,0.3366)
(0.9320,0.3366)

\rput[r](0.1330,0.3366){ 5000}
\PST@Border(0.1490,0.4629)
(0.1640,0.4629)

\PST@Border(0.9470,0.4629)
(0.9320,0.4629)

\rput[r](0.1330,0.4629){ 10000}
\PST@Border(0.1490,0.5891)
(0.1640,0.5891)

\PST@Border(0.9470,0.5891)
(0.9320,0.5891)

\rput(0,0.5891){bits}
\rput[r](0.1330,0.5891){ 15000}
\PST@Border(0.1490,0.7154)
(0.1640,0.7154)

\PST@Border(0.9470,0.7154)
(0.9320,0.7154)

\rput[r](0.1330,0.7154){ 20000}
\PST@Border(0.1490,0.8417)
(0.1640,0.8417)

\PST@Border(0.9470,0.8417)
(0.9320,0.8417)

\rput[r](0.1330,0.8417){ 25000}
\PST@Border(0.1490,0.9680)
(0.1640,0.9680)

\PST@Border(0.9470,0.9680)
(0.9320,0.9680)

\rput[r](0.1330,0.9680){ 30000}
\PST@Border(0.1490,0.0840)
(0.1490,0.1040)

\PST@Border(0.1490,0.9680)
(0.1490,0.9480)

\rput(0.1490,0.0420){ 0}
\PST@Border(0.2630,0.0840)
(0.2630,0.1040)

\PST@Border(0.2630,0.9680)
(0.2630,0.9480)

\rput(0.2630,0.0420){ 1}
\PST@Border(0.3770,0.0840)
(0.3770,0.1040)

\PST@Border(0.3770,0.9680)
(0.3770,0.9480)

\rput(0.3770,0.0420){ 2}
\PST@Border(0.4910,0.0840)
(0.4910,0.1040)

\PST@Border(0.4910,0.9680)
(0.4910,0.9480)

\rput(0.4910,0.0420){ 3}
\PST@Border(0.6050,0.0840)
(0.6050,0.1040)

\PST@Border(0.6050,0.9680)
(0.6050,0.9480)

\rput(0.6050,0.0420){ 4}
\PST@Border(0.7190,0.0840)
(0.7190,0.1040)

\PST@Border(0.7190,0.9680)
(0.7190,0.9480)

\rput(0.7190,0.0420){ 5}
\PST@Border(0.8330,0.0840)
(0.8330,0.1040)

\PST@Border(0.8330,0.9680)
(0.8330,0.9480)

\rput(0.8330,0.0420){ 6}
\PST@Border(0.9470,0.0840)
(0.9470,0.1040)

\PST@Border(0.9470,0.9680)
(0.9470,0.9480)

\rput(0.9470,0.0420){ 7}
\PST@Border(0.1490,0.9680)
(0.1490,0.0840)
(0.9470,0.0840)
(0.9470,0.9680)
(0.1490,0.9680)

\rput[r](0.8200,0.8430){${\cal C}$ -- Havege}
\PST@Solid(0.8360,0.8430)
(0.9150,0.8430)

\PST@Solid(0.1490,0.2104)
(0.1490,0.2104)
(0.1501,0.5556)
(0.1513,0.6913)
(0.1524,0.5922)
(0.1536,0.5521)
(0.1547,0.5233)
(0.1558,0.5005)
(0.1570,0.5472)
(0.1581,0.5444)
(0.1593,0.5651)
(0.1604,0.6231)
(0.1615,0.6187)
(0.1627,0.6189)
(0.1638,0.6819)
(0.1650,0.6557)
(0.1661,0.5429)
(0.1672,0.5128)
(0.1684,0.4696)
(0.1695,0.3911)
(0.1707,0.4233)
(0.1718,0.4281)
(0.1729,0.4258)
(0.1741,0.4357)
(0.1752,0.4408)
(0.1764,0.4532)
(0.1775,0.4306)
(0.1786,0.4026)
(0.1798,0.4043)
(0.1809,0.4079)
(0.1821,0.4309)
(0.1832,0.4543)
(0.1843,0.4695)
(0.1855,0.4572)
(0.1866,0.5013)
(0.1878,0.5352)
(0.1889,0.5120)
(0.1900,0.5082)
(0.1912,0.5300)
(0.1923,0.5157)
(0.1935,0.4565)
(0.1946,0.4293)
(0.1957,0.4525)
(0.1969,0.4502)
(0.1980,0.4081)
(0.1992,0.3770)
(0.2003,0.3590)
(0.2014,0.3521)
(0.2026,0.3234)
(0.2037,0.3142)
(0.2049,0.3143)
(0.2060,0.3125)
(0.2071,0.3220)
(0.2083,0.3403)
(0.2094,0.3400)
(0.2106,0.3300)
(0.2117,0.3471)
(0.2128,0.3370)
(0.2140,0.3332)
(0.2151,0.3473)
(0.2163,0.3438)
(0.2174,0.3348)
(0.2185,0.3369)
(0.2197,0.3350)
(0.2208,0.3207)
(0.2220,0.3133)
(0.2231,0.3225)
(0.2242,0.3338)
(0.2254,0.3438)
(0.2265,0.3520)
(0.2277,0.3524)
(0.2288,0.3500)
(0.2299,0.3326)
(0.2311,0.3307)
(0.2322,0.3470)
(0.2334,0.3515)
(0.2345,0.3416)
(0.2356,0.3330)
(0.2368,0.3350)
(0.2379,0.3539)
(0.2391,0.3511)
(0.2402,0.3363)
(0.2413,0.3188)
(0.2425,0.3097)
(0.2436,0.3280)
(0.2448,0.3473)
(0.2459,0.3633)
(0.2470,0.3514)
(0.2482,0.3473)
(0.2493,0.3869)
(0.2505,0.3929)
(0.2516,0.3826)
(0.2527,0.3729)
(0.2539,0.3662)
(0.2550,0.3732)
(0.2562,0.3780)
(0.2573,0.3820)
(0.2584,0.3874)
(0.2596,0.3807)
(0.2607,0.3583)
(0.2619,0.3498)
\PST@Solid(0.2619,0.3498)
(0.2630,0.3622)
(0.2641,0.3918)
(0.2653,0.4010)
(0.2664,0.3881)
(0.2676,0.3922)
(0.2687,0.3976)
(0.2698,0.4120)
(0.2710,0.4276)
(0.2721,0.4273)
(0.2733,0.4249)
(0.2744,0.4174)
(0.2755,0.4087)
(0.2767,0.4076)
(0.2778,0.4176)
(0.2790,0.4157)
(0.2801,0.4113)
(0.2812,0.4042)
(0.2824,0.4116)
(0.2835,0.4081)
(0.2847,0.4192)
(0.2858,0.4179)
(0.2869,0.4112)
(0.2881,0.4151)
(0.2892,0.4081)
(0.2904,0.4126)
(0.2915,0.4263)
(0.2926,0.4158)
(0.2938,0.4123)
(0.2949,0.4117)
(0.2961,0.4122)
(0.2972,0.4040)
(0.2983,0.4111)
(0.2995,0.4210)
(0.3006,0.4116)
(0.3018,0.4277)
(0.3029,0.4180)
(0.3040,0.4123)
(0.3052,0.4203)
(0.3063,0.4122)
(0.3075,0.3970)
(0.3086,0.3357)
(0.3097,0.3795)
(0.3109,0.4253)
(0.3120,0.4287)
(0.3132,0.4202)
(0.3143,0.4250)
(0.3154,0.4310)
(0.3166,0.4175)
(0.3177,0.4247)
(0.3189,0.4365)
(0.3200,0.4308)
(0.3211,0.4454)
(0.3223,0.4518)
(0.3234,0.4301)
(0.3246,0.4092)
(0.3257,0.3969)
(0.3268,0.3700)
(0.3280,0.3709)
(0.3291,0.3727)
(0.3303,0.3647)
(0.3314,0.3754)
(0.3325,0.4081)
(0.3337,0.4306)
(0.3348,0.4233)
(0.3360,0.4069)
(0.3371,0.4068)
(0.3382,0.4268)
(0.3394,0.4520)
(0.3405,0.4598)
(0.3417,0.4479)
(0.3428,0.4411)
(0.3439,0.4371)
(0.3451,0.4378)
(0.3462,0.4224)
(0.3474,0.3934)
(0.3485,0.3866)
(0.3496,0.4098)
(0.3508,0.4155)
(0.3519,0.4085)
(0.3531,0.4177)
(0.3542,0.4301)
(0.3553,0.4254)
(0.3565,0.4227)
(0.3576,0.4412)
(0.3588,0.4623)
(0.3599,0.4658)
(0.3610,0.4721)
(0.3622,0.4663)
(0.3633,0.4625)
(0.3645,0.4678)
(0.3656,0.4705)
(0.3667,0.4688)
(0.3679,0.4620)
(0.3690,0.4607)
(0.3702,0.4663)
(0.3713,0.4608)
(0.3724,0.4605)
(0.3736,0.4497)
(0.3747,0.4575)
(0.3759,0.4605)
\PST@Solid(0.3759,0.4605)
(0.3770,0.4606)
(0.3781,0.4693)
(0.3793,0.4673)
(0.3804,0.4690)
(0.3816,0.4666)
(0.3827,0.4692)
(0.3838,0.4676)
(0.3850,0.4595)
(0.3861,0.4677)
(0.3873,0.4578)
(0.3884,0.4517)
(0.3895,0.4563)
(0.3907,0.4435)
(0.3918,0.4601)
(0.3930,0.4633)
(0.3941,0.4586)
(0.3952,0.4691)
(0.3964,0.4529)
(0.3975,0.4400)
(0.3987,0.4492)
(0.3998,0.4595)
(0.4009,0.4616)
(0.4021,0.4579)
(0.4032,0.4756)
(0.4044,0.4963)
(0.4055,0.5060)
(0.4066,0.5098)
(0.4078,0.4811)
(0.4089,0.4429)
(0.4101,0.4011)
(0.4112,0.3704)
(0.4123,0.3577)
(0.4135,0.3549)
(0.4146,0.3688)
(0.4158,0.3825)
(0.4169,0.4046)
(0.4180,0.4071)
(0.4192,0.4076)
(0.4203,0.4137)
(0.4215,0.4146)
(0.4226,0.4101)
(0.4237,0.4120)
(0.4249,0.4027)
(0.4260,0.3824)
(0.4272,0.3580)
(0.4283,0.3484)
(0.4294,0.3409)
(0.4306,0.3435)
(0.4317,0.3441)
(0.4329,0.3356)
(0.4340,0.3219)
(0.4351,0.3163)
(0.4363,0.3134)
(0.4374,0.3166)
(0.4386,0.3217)
(0.4397,0.3277)
(0.4408,0.3274)
(0.4420,0.3256)
(0.4431,0.3309)
(0.4443,0.3299)
(0.4454,0.3315)
(0.4465,0.3371)
(0.4477,0.3328)
(0.4488,0.3274)
(0.4500,0.3337)
(0.4511,0.3378)
(0.4522,0.3310)
(0.4534,0.3339)
(0.4545,0.3429)
(0.4557,0.3460)
(0.4568,0.3390)
(0.4579,0.3446)
(0.4591,0.3458)
(0.4602,0.3433)
(0.4614,0.3405)
(0.4625,0.3414)
(0.4636,0.3413)
(0.4648,0.3438)
(0.4659,0.3499)
(0.4671,0.3568)
(0.4682,0.3522)
(0.4693,0.3481)
(0.4705,0.3511)
(0.4716,0.3456)
(0.4728,0.3443)
(0.4739,0.3501)
(0.4750,0.3531)
(0.4762,0.3504)
(0.4773,0.3543)
(0.4785,0.3587)
(0.4796,0.3627)
(0.4807,0.3554)
(0.4819,0.3489)
(0.4830,0.3459)
(0.4842,0.3512)
(0.4853,0.3501)
(0.4864,0.3496)
(0.4876,0.3587)
(0.4887,0.3505)
(0.4899,0.3405)
\PST@Solid(0.4899,0.3405)
(0.4910,0.3399)
(0.4921,0.3422)
(0.4933,0.3484)
(0.4944,0.3461)
(0.4956,0.3437)
(0.4967,0.3563)
(0.4978,0.3590)
(0.4990,0.3510)
(0.5001,0.3554)
(0.5013,0.3514)
(0.5024,0.3461)
(0.5035,0.3385)
(0.5047,0.3350)
(0.5058,0.3363)
(0.5070,0.3308)
(0.5081,0.3300)
(0.5092,0.3532)
(0.5104,0.3621)
(0.5115,0.3320)
(0.5127,0.2946)
(0.5138,0.3156)
(0.5149,0.3314)
(0.5161,0.3282)
(0.5172,0.3334)
(0.5184,0.3401)
(0.5195,0.3696)
(0.5206,0.3760)
(0.5218,0.3585)
(0.5229,0.3290)
(0.5241,0.3257)
(0.5252,0.3279)
(0.5263,0.3410)
(0.5275,0.3496)
(0.5286,0.3513)
(0.5298,0.3494)
(0.5309,0.3466)
(0.5320,0.3431)
(0.5332,0.3523)
(0.5343,0.3538)
(0.5355,0.3534)
(0.5366,0.3513)
(0.5377,0.3499)
(0.5389,0.3408)
(0.5400,0.3439)
(0.5412,0.3446)
(0.5423,0.3486)
(0.5434,0.3553)
(0.5446,0.3460)
(0.5457,0.3446)
(0.5469,0.3537)
(0.5480,0.3512)
(0.5491,0.3453)
(0.5503,0.3452)
(0.5514,0.3646)
(0.5526,0.3701)
(0.5537,0.3731)
(0.5548,0.3713)
(0.5560,0.3578)
(0.5571,0.3568)
(0.5583,0.3529)
(0.5594,0.3505)
(0.5605,0.3436)
(0.5617,0.3399)
(0.5628,0.3365)
(0.5640,0.3418)
(0.5651,0.3456)
(0.5662,0.3447)
(0.5674,0.3442)
(0.5685,0.3476)
(0.5697,0.3420)
(0.5708,0.3393)
(0.5719,0.3467)
(0.5731,0.3515)
(0.5742,0.3455)
(0.5754,0.3456)
(0.5765,0.3393)
(0.5776,0.3394)
(0.5788,0.3427)
(0.5799,0.3387)
(0.5811,0.3399)
(0.5822,0.3385)
(0.5833,0.3414)
(0.5845,0.3405)
(0.5856,0.3384)
(0.5868,0.3486)
(0.5879,0.3562)
(0.5890,0.3446)
(0.5902,0.3406)
(0.5913,0.3385)
(0.5925,0.3325)
(0.5936,0.3360)
(0.5947,0.3352)
(0.5959,0.3378)
(0.5970,0.3418)
(0.5982,0.3398)
(0.5993,0.3322)
(0.6004,0.3366)
(0.6016,0.4049)
(0.6027,0.5247)
(0.6039,0.5918)
\PST@Solid(0.6039,0.5918)
(0.6050,0.6227)
(0.6061,0.6473)
(0.6073,0.6747)
(0.6084,0.6863)
(0.6096,0.6967)
(0.6107,0.7149)
(0.6118,0.6846)
(0.6130,0.7027)
(0.6141,0.7142)
(0.6153,0.7463)
(0.6164,0.7459)
(0.6175,0.7284)
(0.6187,0.7200)
(0.6198,0.7174)
(0.6210,0.7228)
(0.6221,0.7274)
(0.6232,0.7113)
(0.6244,0.7013)
(0.6255,0.7040)
(0.6267,0.7072)
(0.6278,0.6940)
(0.6289,0.6897)
(0.6301,0.7123)
(0.6312,0.7264)
(0.6324,0.7406)
(0.6335,0.7622)
(0.6346,0.7366)
(0.6358,0.7094)
(0.6369,0.6703)
(0.6381,0.6713)
(0.6392,0.7524)
(0.6403,0.8157)
(0.6415,0.8209)
(0.6426,0.8259)
(0.6438,0.8211)
(0.6449,0.8451)
(0.6460,0.8466)
(0.6472,0.8253)
(0.6483,0.8027)
(0.6495,0.8011)
(0.6506,0.8002)
(0.6517,0.7879)
(0.6529,0.8008)
(0.6540,0.8006)
(0.6552,0.7829)
(0.6563,0.8090)
(0.6574,0.8205)
(0.6586,0.7944)
(0.6597,0.7878)
(0.6609,0.7908)
(0.6620,0.7917)
(0.6631,0.8126)
(0.6643,0.8104)
(0.6654,0.7934)
(0.6666,0.8065)
(0.6677,0.8116)
(0.6688,0.7790)
(0.6700,0.7859)
(0.6711,0.8041)
(0.6723,0.7978)
(0.6734,0.7828)
(0.6745,0.7873)
(0.6757,0.7727)
(0.6768,0.7586)
(0.6780,0.7535)
(0.6791,0.7408)
(0.6802,0.7398)
(0.6814,0.7432)
(0.6825,0.7558)
(0.6837,0.7538)
(0.6848,0.7568)
(0.6859,0.7704)
(0.6871,0.7641)
(0.6882,0.7510)
(0.6894,0.7486)
(0.6905,0.7469)
(0.6916,0.7506)
(0.6928,0.7613)
(0.6939,0.7649)
(0.6951,0.7735)
(0.6962,0.7754)
(0.6973,0.7832)
(0.6985,0.7840)
(0.6996,0.7952)
(0.7008,0.7833)
(0.7019,0.7714)
(0.7030,0.7561)
(0.7042,0.7383)
(0.7053,0.7481)
(0.7065,0.7688)
(0.7076,0.7461)
(0.7087,0.7394)
(0.7099,0.7505)
(0.7110,0.7520)
(0.7122,0.7661)
(0.7133,0.7529)
(0.7144,0.7446)
(0.7156,0.7122)
(0.7167,0.7036)
(0.7179,0.6892)
\PST@Solid(0.7179,0.6892)
(0.7190,0.6406)
(0.7201,0.6428)
(0.7213,0.6899)
(0.7224,0.7142)
(0.7236,0.7002)
(0.7247,0.7291)
(0.7258,0.7425)
(0.7270,0.7360)
(0.7281,0.7296)
(0.7293,0.7278)
(0.7304,0.7309)
(0.7315,0.7247)
(0.7327,0.7065)
(0.7338,0.7074)
(0.7350,0.7277)
(0.7361,0.7483)
(0.7372,0.7350)
(0.7384,0.7292)
(0.7395,0.7298)
(0.7407,0.7263)
(0.7418,0.7304)
(0.7429,0.7371)
(0.7441,0.7520)
(0.7452,0.7258)
(0.7464,0.7011)
(0.7475,0.7394)
(0.7486,0.7402)
(0.7498,0.7180)
(0.7509,0.7197)
(0.7521,0.7207)
(0.7532,0.7284)
(0.7543,0.7217)
(0.7555,0.7411)
(0.7566,0.7516)
(0.7578,0.7291)
(0.7589,0.7185)
(0.7600,0.7327)
(0.7612,0.7380)
(0.7623,0.7419)
(0.7635,0.7533)
(0.7646,0.7408)
(0.7657,0.7215)
(0.7669,0.7160)
(0.7680,0.7337)
(0.7692,0.7637)
(0.7703,0.7539)
(0.7714,0.7377)
(0.7726,0.7509)
(0.7737,0.7334)
(0.7749,0.7258)
(0.7760,0.7409)
(0.7771,0.7491)
(0.7783,0.7290)
(0.7794,0.6923)
(0.7806,0.7067)
(0.7817,0.7139)
(0.7828,0.7015)
(0.7840,0.7329)
(0.7851,0.7554)
(0.7863,0.7687)
(0.7874,0.7721)
(0.7885,0.7728)
(0.7897,0.7359)
(0.7908,0.7374)
(0.7920,0.7403)
(0.7931,0.7401)
(0.7942,0.7478)
(0.7954,0.7712)
(0.7965,0.7592)
(0.7977,0.7410)
(0.7988,0.7594)
(0.7999,0.7533)
(0.8011,0.7305)
(0.8022,0.7399)
(0.8034,0.7371)
(0.8045,0.7251)
(0.8056,0.7090)
(0.8068,0.7244)
(0.8079,0.7261)
(0.8091,0.7103)
(0.8102,0.7129)
(0.8113,0.7046)
(0.8125,0.7045)
(0.8136,0.7130)
(0.8148,0.6975)
(0.8159,0.7029)
(0.8170,0.7306)
(0.8182,0.7183)
(0.8193,0.7085)
(0.8205,0.7057)
(0.8216,0.7068)
(0.8227,0.7119)
(0.8239,0.6995)
(0.8250,0.7004)
(0.8262,0.7060)
(0.8273,0.6996)
(0.8284,0.6956)
(0.8296,0.6897)
(0.8307,0.7004)
(0.8319,0.6752)
\PST@Solid(0.8319,0.6752)
(0.8330,0.6726)
(0.8341,0.6743)
(0.8353,0.6801)
(0.8364,0.6805)
(0.8376,0.6609)
(0.8387,0.6582)
(0.8398,0.6545)
(0.8410,0.6489)
(0.8421,0.6434)
(0.8433,0.6425)
(0.8444,0.6120)
(0.8455,0.6108)
(0.8467,0.6305)
(0.8478,0.6279)
(0.8490,0.6005)
(0.8501,0.5742)
(0.8512,0.5726)
(0.8524,0.5580)
(0.8535,0.5426)
(0.8547,0.5257)
(0.8558,0.4789)
(0.8569,0.3982)
(0.8581,0.3246)
(0.8592,0.2882)
(0.8604,0.2648)
(0.8615,0.2372)
(0.8626,0.2306)
(0.8638,0.2280)
(0.8649,0.2240)

\rput[r](0.8200,0.8010){$100\Delta$ -- Havege}
\PST@Dashed(0.8360,0.8010)
(0.9150,0.8010)

\PST@Dashed(0.1490,0.2103)
(0.1490,0.2103)
(0.1501,0.2194)
(0.1513,0.2153)
(0.1524,0.2169)
(0.1536,0.2103)
(0.1547,0.2128)
(0.1558,0.2126)
(0.1570,0.2103)
(0.1581,0.2156)
(0.1593,0.2103)
(0.1604,0.2100)
(0.1615,0.2103)
(0.1627,0.2169)
(0.1638,0.2126)
(0.1650,0.2153)
(0.1661,0.2153)
(0.1672,0.2169)
(0.1684,0.2128)
(0.1695,0.2179)
(0.1707,0.2153)
(0.1718,0.2103)
(0.1729,0.2103)
(0.1741,0.2128)
(0.1752,0.2128)
(0.1764,0.2103)
(0.1775,0.2103)
(0.1786,0.2128)
(0.1798,0.2103)
(0.1809,0.2128)
(0.1821,0.2103)
(0.1832,0.2128)
(0.1843,0.2103)
(0.1855,0.2103)
(0.1866,0.2103)
(0.1878,0.2100)
(0.1889,0.2103)
(0.1900,0.2103)
(0.1912,0.2103)
(0.1923,0.2103)
(0.1935,0.2103)
(0.1946,0.2103)
(0.1957,0.2103)
(0.1969,0.2103)
(0.1980,0.2103)
(0.1992,0.2103)
(0.2003,0.2169)
(0.2014,0.2103)
(0.2026,0.2103)
(0.2037,0.2103)
(0.2049,0.2103)
(0.2060,0.2103)
(0.2071,0.2103)
(0.2083,0.2103)
(0.2094,0.2103)
(0.2106,0.2128)
(0.2117,0.2103)
(0.2128,0.2103)
(0.2140,0.2103)
(0.2151,0.2103)
(0.2163,0.2103)
(0.2174,0.2103)
(0.2185,0.2103)
(0.2197,0.2103)
(0.2208,0.2103)
(0.2220,0.2128)
(0.2231,0.2103)
(0.2242,0.2103)
(0.2254,0.2103)
(0.2265,0.2103)
(0.2277,0.2103)
(0.2288,0.2103)
(0.2299,0.2103)
(0.2311,0.2103)
(0.2322,0.2103)
(0.2334,0.2103)
(0.2345,0.2103)
(0.2356,0.2103)
(0.2368,0.2103)
(0.2379,0.2128)
(0.2391,0.2103)
(0.2402,0.2103)
(0.2413,0.2103)
(0.2425,0.2103)
(0.2436,0.2103)
(0.2448,0.2103)
(0.2459,0.2103)
(0.2470,0.2103)
(0.2482,0.2103)
(0.2493,0.2103)
(0.2505,0.2103)
(0.2516,0.2103)
(0.2527,0.2103)
(0.2539,0.2103)
(0.2550,0.2103)
(0.2562,0.2103)
(0.2573,0.2103)
(0.2584,0.2103)
(0.2596,0.2103)
(0.2607,0.2103)
(0.2619,0.2103)
\PST@Dashed(0.2619,0.2103)
(0.2630,0.2103)
(0.2641,0.2103)
(0.2653,0.2103)
(0.2664,0.2103)
(0.2676,0.2103)
(0.2687,0.2103)
(0.2698,0.2103)
(0.2710,0.2103)
(0.2721,0.2103)
(0.2733,0.2103)
(0.2744,0.2103)
(0.2755,0.2128)
(0.2767,0.2103)
(0.2778,0.2103)
(0.2790,0.2103)
(0.2801,0.2103)
(0.2812,0.2103)
(0.2824,0.2103)
(0.2835,0.2103)
(0.2847,0.2103)
(0.2858,0.2128)
(0.2869,0.2103)
(0.2881,0.2103)
(0.2892,0.2103)
(0.2904,0.2128)
(0.2915,0.2103)
(0.2926,0.2103)
(0.2938,0.2103)
(0.2949,0.2103)
(0.2961,0.2103)
(0.2972,0.2103)
(0.2983,0.2103)
(0.2995,0.2103)
(0.3006,0.2103)
(0.3018,0.2103)
(0.3029,0.2103)
(0.3040,0.2103)
(0.3052,0.2103)
(0.3063,0.2103)
(0.3075,0.2103)
(0.3086,0.2103)
(0.3097,0.2128)
(0.3109,0.2103)
(0.3120,0.2103)
(0.3132,0.2103)
(0.3143,0.2103)
(0.3154,0.2103)
(0.3166,0.2128)
(0.3177,0.2103)
(0.3189,0.2103)
(0.3200,0.2103)
(0.3211,0.2128)
(0.3223,0.2242)
(0.3234,0.2103)
(0.3246,0.2103)
(0.3257,0.2103)
(0.3268,0.2103)
(0.3280,0.2103)
(0.3291,0.2103)
(0.3303,0.2103)
(0.3314,0.2103)
(0.3325,0.2103)
(0.3337,0.2103)
(0.3348,0.2103)
(0.3360,0.2103)
(0.3371,0.2153)
(0.3382,0.2103)
(0.3394,0.2103)
(0.3405,0.2128)
(0.3417,0.2103)
(0.3428,0.2103)
(0.3439,0.2103)
(0.3451,0.2128)
(0.3462,0.2103)
(0.3474,0.2103)
(0.3485,0.2128)
(0.3496,0.2128)
(0.3508,0.2103)
(0.3519,0.2103)
(0.3531,0.2128)
(0.3542,0.2103)
(0.3553,0.2103)
(0.3565,0.2103)
(0.3576,0.2103)
(0.3588,0.2103)
(0.3599,0.2103)
(0.3610,0.2103)
(0.3622,0.2103)
(0.3633,0.2103)
(0.3645,0.2103)
(0.3656,0.2103)
(0.3667,0.2103)
(0.3679,0.2103)
(0.3690,0.2103)
(0.3702,0.2103)
(0.3713,0.2103)
(0.3724,0.2103)
(0.3736,0.2103)
(0.3747,0.2103)
(0.3759,0.2103)
\PST@Dashed(0.3759,0.2103)
(0.3770,0.2128)
(0.3781,0.2166)
(0.3793,0.2103)
(0.3804,0.2103)
(0.3816,0.2103)
(0.3827,0.2128)
(0.3838,0.2103)
(0.3850,0.2128)
(0.3861,0.2103)
(0.3873,0.2103)
(0.3884,0.2103)
(0.3895,0.2103)
(0.3907,0.2103)
(0.3918,0.2128)
(0.3930,0.2103)
(0.3941,0.2103)
(0.3952,0.2103)
(0.3964,0.2103)
(0.3975,0.2103)
(0.3987,0.2103)
(0.3998,0.2128)
(0.4009,0.2103)
(0.4021,0.2103)
(0.4032,0.2103)
(0.4044,0.2103)
(0.4055,0.2128)
(0.4066,0.2103)
(0.4078,0.2103)
(0.4089,0.2128)
(0.4101,0.2128)
(0.4112,0.2103)
(0.4123,0.2169)
(0.4135,0.2219)
(0.4146,0.2103)
(0.4158,0.2103)
(0.4169,0.2103)
(0.4180,0.2153)
(0.4192,0.2103)
(0.4203,0.2153)
(0.4215,0.2103)
(0.4226,0.2103)
(0.4237,0.2103)
(0.4249,0.2103)
(0.4260,0.2103)
(0.4272,0.2153)
(0.4283,0.2169)
(0.4294,0.2128)
(0.4306,0.2153)
(0.4317,0.2169)
(0.4329,0.2128)
(0.4340,0.2103)
(0.4351,0.2128)
(0.4363,0.2103)
(0.4374,0.2103)
(0.4386,0.2219)
(0.4397,0.2128)
(0.4408,0.2128)
(0.4420,0.2103)
(0.4431,0.2103)
(0.4443,0.2103)
(0.4454,0.2128)
(0.4465,0.2128)
(0.4477,0.2128)
(0.4488,0.2103)
(0.4500,0.2103)
(0.4511,0.2103)
(0.4522,0.2103)
(0.4534,0.2128)
(0.4545,0.2128)
(0.4557,0.2169)
(0.4568,0.2103)
(0.4579,0.2103)
(0.4591,0.2128)
(0.4602,0.2128)
(0.4614,0.2103)
(0.4625,0.2103)
(0.4636,0.2128)
(0.4648,0.2103)
(0.4659,0.2219)
(0.4671,0.2103)
(0.4682,0.2103)
(0.4693,0.2103)
(0.4705,0.2103)
(0.4716,0.2103)
(0.4728,0.2103)
(0.4739,0.2103)
(0.4750,0.2103)
(0.4762,0.2103)
(0.4773,0.2103)
(0.4785,0.2128)
(0.4796,0.2103)
(0.4807,0.2103)
(0.4819,0.2153)
(0.4830,0.2103)
(0.4842,0.2103)
(0.4853,0.2103)
(0.4864,0.2128)
(0.4876,0.2103)
(0.4887,0.2103)
(0.4899,0.2103)
\PST@Dashed(0.4899,0.2103)
(0.4910,0.2103)
(0.4921,0.2103)
(0.4933,0.2103)
(0.4944,0.2128)
(0.4956,0.2103)
(0.4967,0.2128)
(0.4978,0.2153)
(0.4990,0.2103)
(0.5001,0.2128)
(0.5013,0.2103)
(0.5024,0.2128)
(0.5035,0.2128)
(0.5047,0.2103)
(0.5058,0.2103)
(0.5070,0.2103)
(0.5081,0.2103)
(0.5092,0.2128)
(0.5104,0.2103)
(0.5115,0.2103)
(0.5127,0.2219)
(0.5138,0.2103)
(0.5149,0.2103)
(0.5161,0.2103)
(0.5172,0.2103)
(0.5184,0.2103)
(0.5195,0.2128)
(0.5206,0.2103)
(0.5218,0.2103)
(0.5229,0.2103)
(0.5241,0.2103)
(0.5252,0.2128)
(0.5263,0.2128)
(0.5275,0.2128)
(0.5286,0.2219)
(0.5298,0.2128)
(0.5309,0.2103)
(0.5320,0.2169)
(0.5332,0.2103)
(0.5343,0.2103)
(0.5355,0.2128)
(0.5366,0.2153)
(0.5377,0.2128)
(0.5389,0.2128)
(0.5400,0.2103)
(0.5412,0.2128)
(0.5423,0.2128)
(0.5434,0.2128)
(0.5446,0.2128)
(0.5457,0.2153)
(0.5469,0.2153)
(0.5480,0.2128)
(0.5491,0.2103)
(0.5503,0.2103)
(0.5514,0.2128)
(0.5526,0.2103)
(0.5537,0.2103)
(0.5548,0.2103)
(0.5560,0.2103)
(0.5571,0.2103)
(0.5583,0.2103)
(0.5594,0.2103)
(0.5605,0.2103)
(0.5617,0.2103)
(0.5628,0.2103)
(0.5640,0.2103)
(0.5651,0.2103)
(0.5662,0.2103)
(0.5674,0.2103)
(0.5685,0.2103)
(0.5697,0.2103)
(0.5708,0.2103)
(0.5719,0.2103)
(0.5731,0.2103)
(0.5742,0.2103)
(0.5754,0.2103)
(0.5765,0.2128)
(0.5776,0.2103)
(0.5788,0.2103)
(0.5799,0.2103)
(0.5811,0.2103)
(0.5822,0.2103)
(0.5833,0.2103)
(0.5845,0.2128)
(0.5856,0.2103)
(0.5868,0.2219)
(0.5879,0.2169)
(0.5890,0.2128)
(0.5902,0.2103)
(0.5913,0.2103)
(0.5925,0.2103)
(0.5936,0.2103)
(0.5947,0.2103)
(0.5959,0.2103)
(0.5970,0.2103)
(0.5982,0.2103)
(0.5993,0.2103)
(0.6004,0.2103)
(0.6016,0.2103)
(0.6027,0.2103)
(0.6039,0.2105)
\PST@Dashed(0.6039,0.2105)
(0.6050,0.2126)
(0.6061,0.2105)
(0.6073,0.2103)
(0.6084,0.2103)
(0.6096,0.2103)
(0.6107,0.2103)
(0.6118,0.2103)
(0.6130,0.2103)
(0.6141,0.2105)
(0.6153,0.2100)
(0.6164,0.2105)
(0.6175,0.2126)
(0.6187,0.2100)
(0.6198,0.2103)
(0.6210,0.2103)
(0.6221,0.2103)
(0.6232,0.2105)
(0.6244,0.2100)
(0.6255,0.2103)
(0.6267,0.2103)
(0.6278,0.2100)
(0.6289,0.2126)
(0.6301,0.2128)
(0.6312,0.2103)
(0.6324,0.2103)
(0.6335,0.2105)
(0.6346,0.2103)
(0.6358,0.2100)
(0.6369,0.2100)
(0.6381,0.2100)
(0.6392,0.2103)
(0.6403,0.2123)
(0.6415,0.2098)
(0.6426,0.2123)
(0.6438,0.2098)
(0.6449,0.2121)
(0.6460,0.2100)
(0.6472,0.2123)
(0.6483,0.2085)
(0.6495,0.2128)
(0.6506,0.2095)
(0.6517,0.2115)
(0.6529,0.2083)
(0.6540,0.2093)
(0.6552,0.2115)
(0.6563,0.2095)
(0.6574,0.2118)
(0.6586,0.2080)
(0.6597,0.2105)
(0.6609,0.2118)
(0.6620,0.2080)
(0.6631,0.2108)
(0.6643,0.2100)
(0.6654,0.2118)
(0.6666,0.2090)
(0.6677,0.2118)
(0.6688,0.2078)
(0.6700,0.2108)
(0.6711,0.2093)
(0.6723,0.2113)
(0.6734,0.2110)
(0.6745,0.2105)
(0.6757,0.2078)
(0.6768,0.2103)
(0.6780,0.2103)
(0.6791,0.2105)
(0.6802,0.2103)
(0.6814,0.2105)
(0.6825,0.2100)
(0.6837,0.2105)
(0.6848,0.2103)
(0.6859,0.2121)
(0.6871,0.2100)
(0.6882,0.2100)
(0.6894,0.2100)
(0.6905,0.2103)
(0.6916,0.2103)
(0.6928,0.2100)
(0.6939,0.2100)
(0.6951,0.2078)
(0.6962,0.2113)
(0.6973,0.2083)
(0.6985,0.2103)
(0.6996,0.2113)
(0.7008,0.2108)
(0.7019,0.2103)
(0.7030,0.2103)
(0.7042,0.2103)
(0.7053,0.2105)
(0.7065,0.2128)
(0.7076,0.2100)
(0.7087,0.2103)
(0.7099,0.2100)
(0.7110,0.2103)
(0.7122,0.2100)
(0.7133,0.2105)
(0.7144,0.2100)
(0.7156,0.2103)
(0.7167,0.2103)
(0.7179,0.2105)
\PST@Dashed(0.7179,0.2105)
(0.7190,0.2103)
(0.7201,0.2103)
(0.7213,0.2103)
(0.7224,0.2103)
(0.7236,0.2100)
(0.7247,0.2103)
(0.7258,0.2100)
(0.7270,0.2100)
(0.7281,0.2105)
(0.7293,0.2100)
(0.7304,0.2103)
(0.7315,0.2105)
(0.7327,0.2103)
(0.7338,0.2103)
(0.7350,0.2103)
(0.7361,0.2105)
(0.7372,0.2105)
(0.7384,0.2105)
(0.7395,0.2105)
(0.7407,0.2100)
(0.7418,0.2128)
(0.7429,0.2100)
(0.7441,0.2100)
(0.7452,0.2105)
(0.7464,0.2100)
(0.7475,0.2103)
(0.7486,0.2105)
(0.7498,0.2103)
(0.7509,0.2100)
(0.7521,0.2103)
(0.7532,0.2100)
(0.7543,0.2103)
(0.7555,0.2103)
(0.7566,0.2103)
(0.7578,0.2103)
(0.7589,0.2128)
(0.7600,0.2105)
(0.7612,0.2103)
(0.7623,0.2100)
(0.7635,0.2105)
(0.7646,0.2103)
(0.7657,0.2103)
(0.7669,0.2100)
(0.7680,0.2105)
(0.7692,0.2105)
(0.7703,0.2105)
(0.7714,0.2103)
(0.7726,0.2105)
(0.7737,0.2105)
(0.7749,0.2105)
(0.7760,0.2105)
(0.7771,0.2100)
(0.7783,0.2103)
(0.7794,0.2103)
(0.7806,0.2103)
(0.7817,0.2105)
(0.7828,0.2100)
(0.7840,0.2105)
(0.7851,0.2100)
(0.7863,0.2088)
(0.7874,0.2121)
(0.7885,0.2123)
(0.7897,0.2105)
(0.7908,0.2103)
(0.7920,0.2100)
(0.7931,0.2105)
(0.7942,0.2103)
(0.7954,0.2121)
(0.7965,0.2128)
(0.7977,0.2100)
(0.7988,0.2100)
(0.7999,0.2105)
(0.8011,0.2105)
(0.8022,0.2100)
(0.8034,0.2100)
(0.8045,0.2100)
(0.8056,0.2103)
(0.8068,0.2103)
(0.8079,0.2105)
(0.8091,0.2100)
(0.8102,0.2100)
(0.8113,0.2105)
(0.8125,0.2105)
(0.8136,0.2100)
(0.8148,0.2103)
(0.8159,0.2100)
(0.8170,0.2100)
(0.8182,0.2126)
(0.8193,0.2105)
(0.8205,0.2105)
(0.8216,0.2100)
(0.8227,0.2105)
(0.8239,0.2105)
(0.8250,0.2105)
(0.8262,0.2103)
(0.8273,0.2100)
(0.8284,0.2103)
(0.8296,0.2105)
(0.8307,0.2100)
(0.8319,0.2100)
\PST@Dashed(0.8319,0.2100)
(0.8330,0.2103)
(0.8341,0.2105)
(0.8353,0.2105)
(0.8364,0.2100)
(0.8376,0.2105)
(0.8387,0.2103)
(0.8398,0.2100)
(0.8410,0.2105)
(0.8421,0.2103)
(0.8433,0.2103)
(0.8444,0.2100)
(0.8455,0.2103)
(0.8467,0.2100)
(0.8478,0.2100)
(0.8490,0.2103)
(0.8501,0.2103)
(0.8512,0.2103)
(0.8524,0.2103)
(0.8535,0.2105)
(0.8547,0.2103)
(0.8558,0.2103)
(0.8569,0.2103)
(0.8581,0.2128)
(0.8592,0.2153)
(0.8604,0.2194)
(0.8615,0.2153)
(0.8626,0.2103)
(0.8638,0.2103)
(0.8649,0.2103)

\rput[r](0.8200,0.9270){${\cal C}$ -- Normal RNG}
\PST@Dotted(0.8360,0.9270)
(0.9150,0.9270)

\PST@Dotted(0.1490,0.2113)
(0.1490,0.2113)
(0.1501,0.3276)
(0.1513,0.3393)
(0.1524,0.3276)
(0.1536,0.3059)
(0.1547,0.3060)
(0.1558,0.3085)
(0.1570,0.2967)
(0.1581,0.2854)
(0.1593,0.2913)
(0.1604,0.2884)
(0.1615,0.2882)
(0.1627,0.2925)
(0.1638,0.2971)
(0.1650,0.2942)
(0.1661,0.2822)
(0.1672,0.2949)
(0.1684,0.2951)
(0.1695,0.2917)
(0.1707,0.2873)
(0.1718,0.2839)
(0.1729,0.2836)
(0.1741,0.2771)
(0.1752,0.2745)
(0.1764,0.2735)
(0.1775,0.2672)
(0.1786,0.2626)
(0.1798,0.2626)
(0.1809,0.2601)
(0.1821,0.2541)
(0.1832,0.2548)
(0.1843,0.2571)
(0.1855,0.2561)
(0.1866,0.2575)
(0.1878,0.2586)
(0.1889,0.2577)
(0.1900,0.2581)
(0.1912,0.2511)
(0.1923,0.2552)
(0.1935,0.2571)
(0.1946,0.2549)
(0.1957,0.2606)
(0.1969,0.2597)
(0.1980,0.2689)
(0.1992,0.2684)
(0.2003,0.2669)
(0.2014,0.2632)
(0.2026,0.2605)
(0.2037,0.2601)
(0.2049,0.2593)
(0.2060,0.2679)
(0.2071,0.2680)
(0.2083,0.2670)
(0.2094,0.2683)
(0.2106,0.2657)
(0.2117,0.2671)
(0.2128,0.2611)
(0.2140,0.2641)
(0.2151,0.2661)
(0.2163,0.2582)
(0.2174,0.2574)
(0.2185,0.2662)
(0.2197,0.2656)
(0.2208,0.2635)
(0.2220,0.2680)
(0.2231,0.2683)
(0.2242,0.2643)
(0.2254,0.2550)
(0.2265,0.2617)
(0.2277,0.2616)
(0.2288,0.2581)
(0.2299,0.2602)
(0.2311,0.2593)
(0.2322,0.2590)
(0.2334,0.2575)
(0.2345,0.2539)
(0.2356,0.2559)
(0.2368,0.2556)
(0.2379,0.2562)
(0.2391,0.2565)
(0.2402,0.2551)
(0.2413,0.2552)
(0.2425,0.2552)
(0.2436,0.2592)
(0.2448,0.2581)
(0.2459,0.2575)
(0.2470,0.2570)
(0.2482,0.2514)
(0.2493,0.2500)
(0.2505,0.2552)
(0.2516,0.2553)
(0.2527,0.2622)
(0.2539,0.2611)
(0.2550,0.2608)
(0.2562,0.2612)
(0.2573,0.2604)
(0.2584,0.2564)
(0.2596,0.2468)
(0.2607,0.2415)
(0.2619,0.2452)
\PST@Dotted(0.2619,0.2452)
(0.2630,0.2505)
(0.2641,0.2539)
(0.2653,0.2566)
(0.2664,0.2577)
(0.2676,0.2542)
(0.2687,0.2546)
(0.2698,0.2561)
(0.2710,0.2583)
(0.2721,0.2619)
(0.2733,0.2621)
(0.2744,0.2581)
(0.2755,0.2563)
(0.2767,0.2569)
(0.2778,0.2553)
(0.2790,0.2554)
(0.2801,0.2560)
(0.2812,0.2578)
(0.2824,0.2572)
(0.2835,0.2557)
(0.2847,0.2548)
(0.2858,0.2554)
(0.2869,0.2609)
(0.2881,0.2609)
(0.2892,0.2564)
(0.2904,0.2505)
(0.2915,0.2550)
(0.2926,0.2553)
(0.2938,0.2530)
(0.2949,0.2498)
(0.2961,0.2532)
(0.2972,0.2569)
(0.2983,0.2568)
(0.2995,0.2575)
(0.3006,0.2542)
(0.3018,0.2568)
(0.3029,0.2579)
(0.3040,0.2560)
(0.3052,0.2495)
(0.3063,0.2498)
(0.3075,0.2496)
(0.3086,0.2477)
(0.3097,0.2469)
(0.3109,0.2476)
(0.3120,0.2488)
(0.3132,0.2511)
(0.3143,0.2499)
(0.3154,0.2499)
(0.3166,0.2480)
(0.3177,0.2480)
(0.3189,0.2499)
(0.3200,0.2499)
(0.3211,0.2540)
(0.3223,0.2539)
(0.3234,0.2625)
(0.3246,0.2709)
(0.3257,0.2608)
(0.3268,0.2502)
(0.3280,0.2458)
(0.3291,0.2431)
(0.3303,0.2441)
(0.3314,0.2523)
(0.3325,0.2531)
(0.3337,0.2547)
(0.3348,0.2567)
(0.3360,0.2586)
(0.3371,0.2635)
(0.3382,0.2573)
(0.3394,0.2428)
(0.3405,0.2292)
(0.3417,0.2302)
(0.3428,0.2297)
(0.3439,0.2290)
(0.3451,0.2318)
(0.3462,0.2329)
(0.3474,0.2318)
(0.3485,0.2320)
(0.3496,0.2348)
(0.3508,0.2334)
(0.3519,0.2327)
(0.3531,0.2336)
(0.3542,0.2352)
(0.3553,0.2344)
(0.3565,0.2331)
(0.3576,0.2336)
(0.3588,0.2380)
(0.3599,0.2366)
(0.3610,0.2326)
(0.3622,0.2328)
(0.3633,0.2332)
(0.3645,0.2330)
(0.3656,0.2370)
(0.3667,0.2414)
(0.3679,0.2400)
(0.3690,0.2334)
(0.3702,0.2302)
(0.3713,0.2343)
(0.3724,0.2460)
(0.3736,0.2540)
(0.3747,0.2559)
(0.3759,0.2551)
\PST@Dotted(0.3759,0.2551)
(0.3770,0.2574)
(0.3781,0.2548)
(0.3793,0.2556)
(0.3804,0.2727)
(0.3816,0.3025)
(0.3827,0.3174)
(0.3838,0.3208)
(0.3850,0.3313)
(0.3861,0.3403)
(0.3873,0.3568)
(0.3884,0.3699)
(0.3895,0.3809)
(0.3907,0.3875)
(0.3918,0.3930)
(0.3930,0.4037)
(0.3941,0.4248)
(0.3952,0.4326)
(0.3964,0.4425)
(0.3975,0.4393)
(0.3987,0.4498)
(0.3998,0.4630)
(0.4009,0.4659)
(0.4021,0.4634)
(0.4032,0.4606)
(0.4044,0.4620)
(0.4055,0.4641)
(0.4066,0.4744)
(0.4078,0.4752)
(0.4089,0.4936)
(0.4101,0.4883)
(0.4112,0.4791)
(0.4123,0.4840)
(0.4135,0.4751)
(0.4146,0.4717)
(0.4158,0.4950)
(0.4169,0.5016)
(0.4180,0.5010)
(0.4192,0.4861)
(0.4203,0.4942)
(0.4215,0.4891)
(0.4226,0.4836)
(0.4237,0.5039)
(0.4249,0.6087)
(0.4260,0.6812)
(0.4272,0.7349)
(0.4283,0.7579)
(0.4294,0.7651)
(0.4306,0.7576)
(0.4317,0.7685)
(0.4329,0.7403)
(0.4340,0.7500)
(0.4351,0.7715)
(0.4363,0.7853)
(0.4374,0.8001)
(0.4386,0.8091)
(0.4397,0.7913)
(0.4408,0.8058)
(0.4420,0.7722)
(0.4431,0.7518)
(0.4443,0.7351)
(0.4454,0.7164)
(0.4465,0.7281)
(0.4477,0.7561)
(0.4488,0.7555)
(0.4500,0.7601)
(0.4511,0.7453)
(0.4522,0.7524)
(0.4534,0.7473)
(0.4545,0.7427)
(0.4557,0.7720)
(0.4568,0.7821)
(0.4579,0.7833)
(0.4591,0.7770)
(0.4602,0.7767)
(0.4614,0.7539)
(0.4625,0.7630)
(0.4636,0.7680)
(0.4648,0.7589)
(0.4659,0.7830)
(0.4671,0.8066)
(0.4682,0.8132)
(0.4693,0.8138)
(0.4705,0.7897)
(0.4716,0.7918)
(0.4728,0.7849)
(0.4739,0.7776)
(0.4750,0.8119)
(0.4762,0.7937)
(0.4773,0.8245)
(0.4785,0.7989)
(0.4796,0.7730)
(0.4807,0.7617)
(0.4819,0.7628)
(0.4830,0.7323)
(0.4842,0.7531)
(0.4853,0.7828)
(0.4864,0.7535)
(0.4876,0.7768)
(0.4887,0.7880)
(0.4899,0.7683)
\PST@Dotted(0.4899,0.7683)
(0.4910,0.7895)
(0.4921,0.7750)
(0.4933,0.7788)
(0.4944,0.7543)
(0.4956,0.7544)
(0.4967,0.7547)
(0.4978,0.7967)
(0.4990,0.7849)
(0.5001,0.7679)
(0.5013,0.7768)
(0.5024,0.7605)
(0.5035,0.7703)
(0.5047,0.7769)
(0.5058,0.7796)
(0.5070,0.7554)
(0.5081,0.7536)
(0.5092,0.7618)
(0.5104,0.7600)
(0.5115,0.7509)
(0.5127,0.7562)
(0.5138,0.7711)
(0.5149,0.7636)
(0.5161,0.7652)
(0.5172,0.7861)
(0.5184,0.7974)
(0.5195,0.7830)
(0.5206,0.7778)
(0.5218,0.7651)
(0.5229,0.7953)
(0.5241,0.7736)
(0.5252,0.7664)
(0.5263,0.7424)
(0.5275,0.7223)
(0.5286,0.7028)
(0.5298,0.7054)
(0.5309,0.7083)
(0.5320,0.7151)
(0.5332,0.7250)
(0.5343,0.7018)
(0.5355,0.7115)
(0.5366,0.7137)
(0.5377,0.7109)
(0.5389,0.7099)
(0.5400,0.6888)
(0.5412,0.7243)
(0.5423,0.7171)
(0.5434,0.7033)
(0.5446,0.7424)
(0.5457,0.7598)
(0.5469,0.7533)
(0.5480,0.7319)
(0.5491,0.7302)
(0.5503,0.7300)
(0.5514,0.7569)
(0.5526,0.7727)
(0.5537,0.7687)
(0.5548,0.7562)
(0.5560,0.7426)
(0.5571,0.7468)
(0.5583,0.7323)
(0.5594,0.7292)
(0.5605,0.7321)
(0.5617,0.7099)
(0.5628,0.7078)
(0.5640,0.7137)
(0.5651,0.7071)
(0.5662,0.7105)
(0.5674,0.7102)
(0.5685,0.7262)
(0.5697,0.7353)
(0.5708,0.7254)
(0.5719,0.7177)
(0.5731,0.7305)
(0.5742,0.7281)
(0.5754,0.7023)
(0.5765,0.7132)
(0.5776,0.6972)
(0.5788,0.6586)
(0.5799,0.6323)
(0.5811,0.6634)
(0.5822,0.6805)
(0.5833,0.6970)
(0.5845,0.7104)
(0.5856,0.6846)
(0.5868,0.6725)
(0.5879,0.6703)
(0.5890,0.6568)
(0.5902,0.6044)
(0.5913,0.5405)
(0.5925,0.4231)
(0.5936,0.2875)
(0.5947,0.2368)
(0.5959,0.2335)
(0.5970,0.2359)
(0.5982,0.2344)
(0.5993,0.2287)
(0.6004,0.2238)
(0.6016,0.2236)
(0.6027,0.2275)
(0.6039,0.2296)
\PST@Dotted(0.6039,0.2296)
(0.6050,0.2285)
(0.6061,0.2249)
(0.6073,0.2214)
(0.6084,0.2235)
(0.6096,0.2243)
(0.6107,0.2193)
(0.6118,0.2181)
(0.6130,0.2171)
(0.6141,0.2204)
(0.6153,0.2204)
(0.6164,0.2175)
(0.6175,0.2168)
(0.6187,0.2172)
(0.6198,0.2195)
(0.6210,0.2220)
(0.6221,0.2209)
(0.6232,0.2176)
(0.6244,0.2177)
(0.6255,0.2177)
(0.6267,0.2178)
(0.6278,0.2189)
(0.6289,0.2156)
(0.6301,0.2152)
(0.6312,0.2154)
(0.6324,0.2164)
(0.6335,0.2166)
(0.6346,0.2171)
(0.6358,0.2152)
(0.6369,0.2145)
(0.6381,0.2166)
(0.6392,0.2155)
(0.6403,0.2165)
(0.6415,0.2175)
(0.6426,0.2166)
(0.6438,0.2142)
(0.6449,0.2140)
(0.6460,0.2138)
(0.6472,0.2133)
(0.6483,0.2149)
(0.6495,0.2160)
(0.6506,0.2142)
(0.6517,0.2148)
(0.6529,0.2145)
(0.6540,0.2122)
(0.6552,0.2116)
(0.6563,0.2110)
(0.6574,0.2107)
(0.6586,0.2114)
(0.6597,0.2126)

\rput[r](0.8200,0.8850){$100\Delta$ -- Normal RNG}
\PST@LongDash(0.8360,0.8850)
(0.9150,0.8850)

\PST@LongDash(0.1490,0.2113)
(0.1490,0.2113)
(0.1501,0.2103)
(0.1513,0.2103)
(0.1524,0.2103)
(0.1536,0.2103)
(0.1547,0.2128)
(0.1558,0.2103)
(0.1570,0.2103)
(0.1581,0.2103)
(0.1593,0.2103)
(0.1604,0.2103)
(0.1615,0.2128)
(0.1627,0.2103)
(0.1638,0.2103)
(0.1650,0.2103)
(0.1661,0.2103)
(0.1672,0.2103)
(0.1684,0.2103)
(0.1695,0.2103)
(0.1707,0.2103)
(0.1718,0.2103)
(0.1729,0.2103)
(0.1741,0.2103)
(0.1752,0.2103)
(0.1764,0.2103)
(0.1775,0.2103)
(0.1786,0.2103)
(0.1798,0.2103)
(0.1809,0.2103)
(0.1821,0.2103)
(0.1832,0.2103)
(0.1843,0.2103)
(0.1855,0.2103)
(0.1866,0.2103)
(0.1878,0.2103)
(0.1889,0.2103)
(0.1900,0.2103)
(0.1912,0.2103)
(0.1923,0.2103)
(0.1935,0.2103)
(0.1946,0.2103)
(0.1957,0.2103)
(0.1969,0.2103)
(0.1980,0.2103)
(0.1992,0.2103)
(0.2003,0.2103)
(0.2014,0.2103)
(0.2026,0.2103)
(0.2037,0.2103)
(0.2049,0.2103)
(0.2060,0.2103)
(0.2071,0.2103)
(0.2083,0.2103)
(0.2094,0.2128)
(0.2106,0.2103)
(0.2117,0.2103)
(0.2128,0.2103)
(0.2140,0.2103)
(0.2151,0.2103)
(0.2163,0.2103)
(0.2174,0.2103)
(0.2185,0.2103)
(0.2197,0.2103)
(0.2208,0.2103)
(0.2220,0.2103)
(0.2231,0.2103)
(0.2242,0.2103)
(0.2254,0.2103)
(0.2265,0.2103)
(0.2277,0.2103)
(0.2288,0.2103)
(0.2299,0.2103)
(0.2311,0.2103)
(0.2322,0.2103)
(0.2334,0.2103)
(0.2345,0.2103)
(0.2356,0.2103)
(0.2368,0.2103)
(0.2379,0.2103)
(0.2391,0.2103)
(0.2402,0.2103)
(0.2413,0.2103)
(0.2425,0.2103)
(0.2436,0.2103)
(0.2448,0.2103)
(0.2459,0.2103)
(0.2470,0.2103)
(0.2482,0.2103)
(0.2493,0.2103)
(0.2505,0.2103)
(0.2516,0.2103)
(0.2527,0.2103)
(0.2539,0.2103)
(0.2550,0.2103)
(0.2562,0.2103)
(0.2573,0.2103)
(0.2584,0.2103)
(0.2596,0.2169)
(0.2607,0.2386)
(0.2619,0.2219)
\PST@LongDash(0.2619,0.2219)
(0.2630,0.2103)
(0.2641,0.2103)
(0.2653,0.2103)
(0.2664,0.2103)
(0.2676,0.2103)
(0.2687,0.2103)
(0.2698,0.2103)
(0.2710,0.2103)
(0.2721,0.2103)
(0.2733,0.2103)
(0.2744,0.2103)
(0.2755,0.2103)
(0.2767,0.2103)
(0.2778,0.2103)
(0.2790,0.2103)
(0.2801,0.2103)
(0.2812,0.2103)
(0.2824,0.2103)
(0.2835,0.2103)
(0.2847,0.2103)
(0.2858,0.2103)
(0.2869,0.2103)
(0.2881,0.2103)
(0.2892,0.2103)
(0.2904,0.2103)
(0.2915,0.2103)
(0.2926,0.2103)
(0.2938,0.2103)
(0.2949,0.2103)
(0.2961,0.2103)
(0.2972,0.2103)
(0.2983,0.2103)
(0.2995,0.2103)
(0.3006,0.2103)
(0.3018,0.2103)
(0.3029,0.2103)
(0.3040,0.2103)
(0.3052,0.2103)
(0.3063,0.2103)
(0.3075,0.2103)
(0.3086,0.2103)
(0.3097,0.2103)
(0.3109,0.2103)
(0.3120,0.2103)
(0.3132,0.2103)
(0.3143,0.2103)
(0.3154,0.2103)
(0.3166,0.2103)
(0.3177,0.2103)
(0.3189,0.2103)
(0.3200,0.2103)
(0.3211,0.2103)
(0.3223,0.2103)
(0.3234,0.2103)
(0.3246,0.2103)
(0.3257,0.2103)
(0.3268,0.2103)
(0.3280,0.2103)
(0.3291,0.2103)
(0.3303,0.2103)
(0.3314,0.2103)
(0.3325,0.2103)
(0.3337,0.2103)
(0.3348,0.2103)
(0.3360,0.2103)
(0.3371,0.2103)
(0.3382,0.2103)
(0.3394,0.2128)
(0.3405,0.2103)
(0.3417,0.2128)
(0.3428,0.2128)
(0.3439,0.2128)
(0.3451,0.2103)
(0.3462,0.2103)
(0.3474,0.2128)
(0.3485,0.2128)
(0.3496,0.2103)
(0.3508,0.2103)
(0.3519,0.2128)
(0.3531,0.2128)
(0.3542,0.2103)
(0.3553,0.2128)
(0.3565,0.2128)
(0.3576,0.2128)
(0.3588,0.2169)
(0.3599,0.2169)
(0.3610,0.2169)
(0.3622,0.2194)
(0.3633,0.2153)
(0.3645,0.2128)
(0.3656,0.2103)
(0.3667,0.2128)
(0.3679,0.2153)
(0.3690,0.2103)
(0.3702,0.2204)
(0.3713,0.2153)
(0.3724,0.2169)
(0.3736,0.2128)
(0.3747,0.2103)
(0.3759,0.2103)
\PST@LongDash(0.3759,0.2103)
(0.3770,0.2128)
(0.3781,0.2103)
(0.3793,0.2103)
(0.3804,0.2103)
(0.3816,0.2103)
(0.3827,0.2103)
(0.3838,0.2103)
(0.3850,0.2103)
(0.3861,0.2103)
(0.3873,0.2103)
(0.3884,0.2103)
(0.3895,0.2103)
(0.3907,0.2103)
(0.3918,0.2103)
(0.3930,0.2103)
(0.3941,0.2103)
(0.3952,0.2103)
(0.3964,0.2103)
(0.3975,0.2103)
(0.3987,0.2103)
(0.3998,0.2103)
(0.4009,0.2103)
(0.4021,0.2103)
(0.4032,0.2103)
(0.4044,0.2103)
(0.4055,0.2103)
(0.4066,0.2126)
(0.4078,0.2103)
(0.4089,0.2105)
(0.4101,0.2103)
(0.4112,0.2103)
(0.4123,0.2128)
(0.4135,0.2103)
(0.4146,0.2128)
(0.4158,0.2191)
(0.4169,0.2103)
(0.4180,0.2100)
(0.4192,0.2128)
(0.4203,0.2128)
(0.4215,0.2153)
(0.4226,0.2128)
(0.4237,0.2103)
(0.4249,0.2105)
(0.4260,0.2103)
(0.4272,0.2100)
(0.4283,0.2105)
(0.4294,0.2100)
(0.4306,0.2105)
(0.4317,0.2093)
(0.4329,0.2100)
(0.4340,0.2103)
(0.4351,0.2128)
(0.4363,0.2103)
(0.4374,0.2088)
(0.4386,0.2088)
(0.4397,0.2105)
(0.4408,0.2090)
(0.4420,0.2115)
(0.4431,0.2103)
(0.4443,0.2100)
(0.4454,0.2103)
(0.4465,0.2105)
(0.4477,0.2105)
(0.4488,0.2103)
(0.4500,0.2103)
(0.4511,0.2103)
(0.4522,0.2100)
(0.4534,0.2100)
(0.4545,0.2103)
(0.4557,0.2131)
(0.4568,0.2118)
(0.4579,0.2085)
(0.4591,0.2103)
(0.4602,0.2105)
(0.4614,0.2100)
(0.4625,0.2103)
(0.4636,0.2095)
(0.4648,0.2103)
(0.4659,0.2085)
(0.4671,0.2090)
(0.4682,0.2128)
(0.4693,0.2108)
(0.4705,0.2110)
(0.4716,0.2080)
(0.4728,0.2085)
(0.4739,0.2108)
(0.4750,0.2103)
(0.4762,0.2118)
(0.4773,0.2128)
(0.4785,0.2105)
(0.4796,0.2090)
(0.4807,0.2100)
(0.4819,0.2105)
(0.4830,0.2103)
(0.4842,0.2100)
(0.4853,0.2098)
(0.4864,0.2103)
(0.4876,0.2100)
(0.4887,0.2115)
(0.4899,0.2083)
\PST@LongDash(0.4899,0.2083)
(0.4910,0.2085)
(0.4921,0.2113)
(0.4933,0.2090)
(0.4944,0.2100)
(0.4956,0.2103)
(0.4967,0.2105)
(0.4978,0.2105)
(0.4990,0.2105)
(0.5001,0.2121)
(0.5013,0.2100)
(0.5024,0.2105)
(0.5035,0.2121)
(0.5047,0.2075)
(0.5058,0.2115)
(0.5070,0.2103)
(0.5081,0.2100)
(0.5092,0.2100)
(0.5104,0.2103)
(0.5115,0.2100)
(0.5127,0.2105)
(0.5138,0.2085)
(0.5149,0.2105)
(0.5161,0.2100)
(0.5172,0.2118)
(0.5184,0.2103)
(0.5195,0.2078)
(0.5206,0.2098)
(0.5218,0.2100)
(0.5229,0.2115)
(0.5241,0.2088)
(0.5252,0.2103)
(0.5263,0.2105)
(0.5275,0.2100)
(0.5286,0.2103)
(0.5298,0.2100)
(0.5309,0.2103)
(0.5320,0.2100)
(0.5332,0.2103)
(0.5343,0.2105)
(0.5355,0.2103)
(0.5366,0.2100)
(0.5377,0.2100)
(0.5389,0.2103)
(0.5400,0.2100)
(0.5412,0.2103)
(0.5423,0.2103)
(0.5434,0.2103)
(0.5446,0.2100)
(0.5457,0.2103)
(0.5469,0.2105)
(0.5480,0.2100)
(0.5491,0.2103)
(0.5503,0.2103)
(0.5514,0.2103)
(0.5526,0.2088)
(0.5537,0.2118)
(0.5548,0.2103)
(0.5560,0.2103)
(0.5571,0.2103)
(0.5583,0.2103)
(0.5594,0.2103)
(0.5605,0.2105)
(0.5617,0.2100)
(0.5628,0.2103)
(0.5640,0.2100)
(0.5651,0.2105)
(0.5662,0.2105)
(0.5674,0.2103)
(0.5685,0.2103)
(0.5697,0.2105)
(0.5708,0.2105)
(0.5719,0.2103)
(0.5731,0.2100)
(0.5742,0.2105)
(0.5754,0.2103)
(0.5765,0.2105)
(0.5776,0.2103)
(0.5788,0.2103)
(0.5799,0.2103)
(0.5811,0.2105)
(0.5822,0.2103)
(0.5833,0.2103)
(0.5845,0.2103)
(0.5856,0.2105)
(0.5868,0.2103)
(0.5879,0.2103)
(0.5890,0.2103)
(0.5902,0.2103)
(0.5913,0.2128)
(0.5925,0.2103)
(0.5936,0.2128)
(0.5947,0.2270)
(0.5959,0.2204)
(0.5970,0.2219)
(0.5982,0.2128)
(0.5993,0.2179)
(0.6004,0.2169)
(0.6016,0.2103)
(0.6027,0.2103)
(0.6039,0.2103)
\PST@LongDash(0.6039,0.2103)
(0.6050,0.2103)
(0.6061,0.2103)
(0.6073,0.2103)
(0.6084,0.2128)
(0.6096,0.2103)
(0.6107,0.2153)
(0.6118,0.2128)
(0.6130,0.2103)
(0.6141,0.2103)
(0.6153,0.2128)
(0.6164,0.2103)
(0.6175,0.2103)
(0.6187,0.2103)
(0.6198,0.2103)
(0.6210,0.2103)
(0.6221,0.2103)
(0.6232,0.2103)
(0.6244,0.2128)
(0.6255,0.2128)
(0.6267,0.2103)
(0.6278,0.2103)
(0.6289,0.2103)
(0.6301,0.2128)
(0.6312,0.2128)
(0.6324,0.2128)
(0.6335,0.2103)
(0.6346,0.2128)
(0.6358,0.2169)
(0.6369,0.2128)
(0.6381,0.2103)
(0.6392,0.2103)
(0.6403,0.2103)
(0.6415,0.2103)
(0.6426,0.2103)
(0.6438,0.2103)
(0.6449,0.2103)
(0.6460,0.2103)
(0.6472,0.2126)
(0.6483,0.2103)
(0.6495,0.2103)
(0.6506,0.2169)
(0.6517,0.2103)
(0.6529,0.2103)
(0.6540,0.2118)
(0.6552,0.2110)
(0.6563,0.2133)
(0.6574,0.2110)
(0.6586,0.2143)
(0.6597,0.2103)

\PST@Border(0.1490,0.9680)
(0.1490,0.0840)
(0.9470,0.0840)
(0.9470,0.9680)
(0.1490,0.9680)

\catcode`@=12
\fi
\endpspicture

%% file: SlicePow1.tex
\ifx\PSTloaded\undefined
\def\PSTloaded{t}
\psset{arrowsize=.01 3.2 1.4 .3}
\psset{dotsize=.01}
\catcode`@=11

\newpsobject{PST@Border}{psline}{linewidth=.0015,linestyle=solid}
\newpsobject{PST@Axes}{psline}{linewidth=.0015,linestyle=dotted,dotsep=.004}
\newpsobject{PST@Solid}{psline}{linewidth=.0015,linestyle=solid}
\newpsobject{PST@Dashed}{psline}{linewidth=.0015,linestyle=dashed,dash=.01 .01}
\newpsobject{PST@Dotted}{psline}{linewidth=.0025,linestyle=dotted,dotsep=.008}
\newpsobject{PST@LongDash}{psline}{linewidth=.0015,linestyle=dashed,dash=.02 .01}
\newpsobject{PST@Diamond}{psdots}{linewidth=.001,linestyle=solid,dotstyle=square,dotangle=45}
\newpsobject{PST@Filldiamond}{psdots}{linewidth=.001,linestyle=solid,dotstyle=square*,dotangle=45}
\newpsobject{PST@Cross}{psdots}{linewidth=.001,linestyle=solid,dotstyle=+,dotangle=45}
\newpsobject{PST@Plus}{psdots}{linewidth=.001,linestyle=solid,dotstyle=+}
\newpsobject{PST@Square}{psdots}{linewidth=.001,linestyle=solid,dotstyle=square}
\newpsobject{PST@Circle}{psdots}{linewidth=.001,linestyle=solid,dotstyle=o}
\newpsobject{PST@Triangle}{psdots}{linewidth=.001,linestyle=solid,dotstyle=triangle}
\newpsobject{PST@Pentagon}{psdots}{linewidth=.001,linestyle=solid,dotstyle=pentagon}
\newpsobject{PST@Fillsquare}{psdots}{linewidth=.001,linestyle=solid,dotstyle=square*}
\newpsobject{PST@Fillcircle}{psdots}{linewidth=.001,linestyle=solid,dotstyle=*}
\newpsobject{PST@Filltriangle}{psdots}{linewidth=.001,linestyle=solid,dotstyle=triangle*}
\newpsobject{PST@Fillpentagon}{psdots}{linewidth=.001,linestyle=solid,dotstyle=pentagon*}
\newpsobject{PST@Arrow}{psline}{linewidth=.001,linestyle=solid}
\catcode`@=12

\fi
\psset{unit=5.0in,xunit=5.0in,yunit=3.0in}
\pspicture(0.000000,0.000000)(1.000000,1.000000)
\ifx\nofigs\undefined
\catcode`@=11

\PST@Border(0.1490,0.0840)
(0.1640,0.0840)

\PST@Border(0.9470,0.0840)
(0.9320,0.0840)

\rput[r](0.1330,0.0840){-5000}
\PST@Border(0.1490,0.2103)
(0.1640,0.2103)

\PST@Border(0.9470,0.2103)
(0.9320,0.2103)

\rput[r](0.1330,0.2103){ 0}
\PST@Border(0.1490,0.3366)
(0.1640,0.3366)

\PST@Border(0.9470,0.3366)
(0.9320,0.3366)

\rput[r](0.1330,0.3366){ 5000}
\PST@Border(0.1490,0.4629)
(0.1640,0.4629)

\PST@Border(0.9470,0.4629)
(0.9320,0.4629)

\rput[r](0.1330,0.4629){ 10000}
\PST@Border(0.1490,0.5891)
(0.1640,0.5891)

\PST@Border(0.9470,0.5891)
(0.9320,0.5891)

\rput(0,0.5891){bits}
\rput[r](0.1330,0.5891){ 15000}
\PST@Border(0.1490,0.7154)
(0.1640,0.7154)

\PST@Border(0.9470,0.7154)
(0.9320,0.7154)

\rput[r](0.1330,0.7154){ 20000}
\PST@Border(0.1490,0.8417)
(0.1640,0.8417)

\PST@Border(0.9470,0.8417)
(0.9320,0.8417)

\rput[r](0.1330,0.8417){ 25000}
\PST@Border(0.1490,0.9680)
(0.1640,0.9680)

\PST@Border(0.9470,0.9680)
(0.9320,0.9680)

\rput[r](0.1330,0.9680){ 30000}
\PST@Border(0.1490,0.0840)
(0.1490,0.1040)

\PST@Border(0.1490,0.9680)
(0.1490,0.9480)

\rput(0.1490,0.0420){ 0}
\PST@Border(0.2377,0.0840)
(0.2377,0.1040)

\PST@Border(0.2377,0.9680)
(0.2377,0.9480)

\rput(0.2377,0.0420){ 1}
\PST@Border(0.3263,0.0840)
(0.3263,0.1040)

\PST@Border(0.3263,0.9680)
(0.3263,0.9480)

\rput(0.3263,0.0420){ 2}
\PST@Border(0.4150,0.0840)
(0.4150,0.1040)

\PST@Border(0.4150,0.9680)
(0.4150,0.9480)

\rput(0.4150,0.0420){ 3}
\PST@Border(0.5037,0.0840)
(0.5037,0.1040)

\PST@Border(0.5037,0.9680)
(0.5037,0.9480)

\rput(0.5037,0.0420){ 4}
\PST@Border(0.5923,0.0840)
(0.5923,0.1040)

\PST@Border(0.5923,0.9680)
(0.5923,0.9480)

\rput(0.5923,0.0420){ 5}
\PST@Border(0.6810,0.0840)
(0.6810,0.1040)

\PST@Border(0.6810,0.9680)
(0.6810,0.9480)

\rput(0.6810,0.0420){ 6}
\PST@Border(0.7697,0.0840)
(0.7697,0.1040)

\PST@Border(0.7697,0.9680)
(0.7697,0.9480)

\rput(0.7697,0.0420){ 7}
\PST@Border(0.8583,0.0840)
(0.8583,0.1040)

\PST@Border(0.8583,0.9680)
(0.8583,0.9480)

\rput(0.8583,0.0420){ 8}
\PST@Border(0.9470,0.0840)
(0.9470,0.1040)

\PST@Border(0.9470,0.9680)
(0.9470,0.9480)

\rput(0.9470,0.0420){ 9}
\PST@Border(0.1490,0.9680)
(0.1490,0.0840)
(0.9470,0.0840)
(0.9470,0.9680)
(0.1490,0.9680)

\rput[r](0.8200,0.8430){${\cal C}$ -- Havege}
\PST@Solid(0.8360,0.8430)
(0.9150,0.8430)

\PST@Solid(0.1490,0.2111)
(0.1490,0.2111)
(0.1499,0.3890)
(0.1508,0.4472)
(0.1517,0.3382)
(0.1525,0.2537)
(0.1534,0.2790)
(0.1543,0.2962)
(0.1552,0.2922)
(0.1561,0.2819)
(0.1570,0.2957)
(0.1579,0.3096)
(0.1588,0.3277)
(0.1596,0.3405)
(0.1605,0.3175)
(0.1614,0.3235)
(0.1623,0.3541)
(0.1632,0.3513)
(0.1641,0.3489)
(0.1650,0.3362)
(0.1658,0.3352)
(0.1667,0.3452)
(0.1676,0.3337)
(0.1685,0.3085)
(0.1694,0.3241)
(0.1703,0.3422)
(0.1712,0.3416)
(0.1721,0.3125)
(0.1729,0.3163)
(0.1738,0.3020)
(0.1747,0.2565)
(0.1756,0.2583)
(0.1765,0.2872)
(0.1774,0.3174)
(0.1783,0.3110)
(0.1791,0.2792)
(0.1800,0.2635)
(0.1809,0.2774)
(0.1818,0.3141)
(0.1827,0.3326)
(0.1836,0.3254)
(0.1845,0.2983)
(0.1854,0.2699)
(0.1862,0.2813)
(0.1871,0.2827)
(0.1880,0.2740)
(0.1889,0.2863)
(0.1898,0.2947)
(0.1907,0.3213)
(0.1916,0.3303)
(0.1924,0.2806)
(0.1933,0.2703)
(0.1942,0.2773)
(0.1951,0.2838)
(0.1960,0.2939)
(0.1969,0.3618)
(0.1978,0.3530)
(0.1987,0.2682)
(0.1995,0.2624)
(0.2004,0.2777)
(0.2013,0.2969)
(0.2022,0.3783)
(0.2031,0.4263)
(0.2040,0.3797)
(0.2049,0.3147)
(0.2057,0.2764)
(0.2066,0.2508)
(0.2075,0.2829)
(0.2084,0.3265)
(0.2093,0.3298)
(0.2102,0.3225)
(0.2111,0.3119)
(0.2120,0.2868)
(0.2128,0.2737)
(0.2137,0.2714)
(0.2146,0.2943)
(0.2155,0.3021)
(0.2164,0.2818)
(0.2173,0.2728)
(0.2182,0.3032)
(0.2190,0.2907)
(0.2199,0.3004)
(0.2208,0.3407)
(0.2217,0.3406)
(0.2226,0.3180)
(0.2235,0.2967)
(0.2244,0.2961)
(0.2253,0.3040)
(0.2261,0.3204)
(0.2270,0.3156)
(0.2279,0.3041)
(0.2288,0.3098)
(0.2297,0.3283)
(0.2306,0.3311)
(0.2315,0.2762)
(0.2323,0.2380)
(0.2332,0.2473)
(0.2341,0.2478)
(0.2350,0.2642)
(0.2359,0.2882)
(0.2368,0.3158)
\PST@Solid(0.2368,0.3158)
(0.2377,0.3082)
(0.2386,0.2637)
(0.2394,0.2879)
(0.2403,0.3250)
(0.2412,0.3294)
(0.2421,0.3220)
(0.2430,0.2917)
(0.2439,0.2896)
(0.2448,0.2976)
(0.2456,0.2685)
(0.2465,0.2380)
(0.2474,0.2475)
(0.2483,0.2752)
(0.2492,0.2798)
(0.2501,0.2885)
(0.2510,0.2850)
(0.2519,0.2492)
(0.2527,0.2254)
(0.2536,0.2385)
(0.2545,0.2563)
(0.2554,0.2394)
(0.2563,0.2167)
(0.2572,0.2196)
(0.2581,0.2215)
(0.2589,0.2214)
(0.2598,0.3566)
(0.2607,0.4447)
(0.2616,0.4168)
(0.2625,0.3389)
(0.2634,0.2854)
(0.2643,0.2873)
(0.2652,0.2886)
(0.2660,0.2712)
(0.2669,0.2535)
(0.2678,0.2428)
(0.2687,0.2386)
(0.2696,0.3122)
(0.2705,0.3373)
(0.2714,0.3285)
(0.2722,0.3414)
(0.2731,0.3148)
(0.2740,0.2405)
(0.2749,0.2147)
(0.2758,0.2153)
(0.2767,0.2203)
(0.2776,0.2384)
(0.2785,0.2810)
(0.2793,0.3218)
(0.2802,0.3392)
(0.2811,0.3326)
(0.2820,0.3121)
(0.2829,0.2832)
(0.2838,0.2665)
(0.2847,0.3970)
(0.2855,0.4561)
(0.2864,0.4658)
(0.2873,0.4759)
(0.2882,0.4384)
(0.2891,0.4468)
(0.2900,0.4554)
(0.2909,0.4574)
(0.2918,0.4700)
(0.2926,0.4709)
(0.2935,0.4701)
(0.2944,0.4756)
(0.2953,0.4778)
(0.2962,0.4711)
(0.2971,0.4544)
(0.2980,0.4598)
(0.2988,0.4889)
(0.2997,0.5293)
(0.3006,0.5491)
(0.3015,0.5609)
(0.3024,0.5441)
(0.3033,0.4917)
(0.3042,0.4706)
(0.3051,0.4541)
(0.3059,0.4497)
(0.3068,0.4577)
(0.3077,0.4527)
(0.3086,0.4507)
(0.3095,0.4213)
(0.3104,0.4007)
(0.3113,0.4054)
(0.3121,0.4313)
(0.3130,0.4530)
(0.3139,0.4513)
(0.3148,0.4394)
(0.3157,0.4253)
(0.3166,0.4096)
(0.3175,0.3906)
(0.3184,0.3785)
(0.3192,0.3523)
(0.3201,0.3463)
(0.3210,0.3174)
(0.3219,0.2909)
(0.3228,0.3002)
(0.3237,0.3198)
(0.3246,0.3354)
(0.3254,0.3338)
\PST@Solid(0.3254,0.3338)
(0.3263,0.3611)
(0.3272,0.3886)
(0.3281,0.3658)
(0.3290,0.3279)
(0.3299,0.3439)
(0.3308,0.3891)
(0.3317,0.4240)
(0.3325,0.3919)
(0.3334,0.3567)
(0.3343,0.3707)
(0.3352,0.3626)
(0.3361,0.3428)
(0.3370,0.3501)
(0.3379,0.3447)
(0.3387,0.3449)
(0.3396,0.3306)
(0.3405,0.3248)
(0.3414,0.3295)
(0.3423,0.3209)
(0.3432,0.3330)
(0.3441,0.3414)
(0.3450,0.3235)
(0.3458,0.3230)
(0.3467,0.3235)
(0.3476,0.3168)
(0.3485,0.3270)
(0.3494,0.3404)
(0.3503,0.3357)
(0.3512,0.3273)
(0.3520,0.3396)
(0.3529,0.3476)
(0.3538,0.3541)
(0.3547,0.3715)
(0.3556,0.3732)
(0.3565,0.3458)
(0.3574,0.3391)
(0.3583,0.3484)
(0.3591,0.3664)
(0.3600,0.3641)
(0.3609,0.3597)
(0.3618,0.3772)
(0.3627,0.3811)
(0.3636,0.3619)
(0.3645,0.3317)
(0.3653,0.3400)
(0.3662,0.3386)
(0.3671,0.3253)
(0.3680,0.3460)
(0.3689,0.3525)
(0.3698,0.3408)
(0.3707,0.3574)
(0.3716,0.3559)
(0.3724,0.3327)
(0.3733,0.3404)
(0.3742,0.3512)
(0.3751,0.3456)
(0.3760,0.3380)
(0.3769,0.3453)
(0.3778,0.3480)
(0.3786,0.3355)
(0.3795,0.3354)
(0.3804,0.3450)
(0.3813,0.3346)
(0.3822,0.3195)
(0.3831,0.3228)
(0.3840,0.3298)
(0.3849,0.3243)
(0.3857,0.3231)
(0.3866,0.3222)
(0.3875,0.3127)
(0.3884,0.3045)
(0.3893,0.3133)
(0.3902,0.3157)
(0.3911,0.3074)
(0.3919,0.3063)
(0.3928,0.3245)
(0.3937,0.3167)
(0.3946,0.3112)
(0.3955,0.3069)
(0.3964,0.3021)
(0.3973,0.2937)
(0.3982,0.2886)
(0.3990,0.2864)
(0.3999,0.2884)
(0.4008,0.2949)
(0.4017,0.2892)
(0.4026,0.2864)
(0.4035,0.2862)
(0.4044,0.2918)
(0.4052,0.2954)
(0.4061,0.2897)
(0.4070,0.2790)
(0.4079,0.2655)
(0.4088,0.2526)
(0.4097,0.2389)
(0.4106,0.2426)
(0.4115,0.2476)
(0.4123,0.2458)
(0.4132,0.2485)
(0.4141,0.2519)
\PST@Solid(0.4141,0.2519)
(0.4150,0.2560)
(0.4159,0.2651)
(0.4168,0.2682)
(0.4177,0.2729)
(0.4185,0.2755)
(0.4194,0.2842)
(0.4203,0.2959)
(0.4212,0.3128)
(0.4221,0.3156)
(0.4230,0.3278)
(0.4239,0.3374)
(0.4248,0.3355)
(0.4256,0.3483)
(0.4265,0.3624)
(0.4274,0.3632)
(0.4283,0.3563)
(0.4292,0.3610)
(0.4301,0.3602)
(0.4310,0.3628)
(0.4318,0.3599)
(0.4327,0.3699)
(0.4336,0.3691)
(0.4345,0.3730)
(0.4354,0.3760)
(0.4363,0.3755)
(0.4372,0.3703)
(0.4381,0.3725)
(0.4389,0.3757)
(0.4398,0.3798)
(0.4407,0.3851)
(0.4416,0.3856)
(0.4425,0.3845)
(0.4434,0.3858)
(0.4443,0.3948)
(0.4451,0.4038)
(0.4460,0.4070)
(0.4469,0.4522)
(0.4478,0.5660)
(0.4487,0.6621)
(0.4496,0.7103)
(0.4505,0.7350)
(0.4514,0.7874)
(0.4522,0.8016)
(0.4531,0.7759)
(0.4540,0.7601)
(0.4549,0.7899)
(0.4558,0.7972)
(0.4567,0.7967)
(0.4576,0.7971)
(0.4584,0.7821)
(0.4593,0.7785)
(0.4602,0.7794)
(0.4611,0.8056)
(0.4620,0.7977)
(0.4629,0.7993)
(0.4638,0.7911)
(0.4647,0.8031)
(0.4655,0.7742)
(0.4664,0.7297)
(0.4673,0.7432)
(0.4682,0.7511)
(0.4691,0.7126)
(0.4700,0.7301)
(0.4709,0.7425)
(0.4717,0.7436)
(0.4726,0.7785)
(0.4735,0.8146)
(0.4744,0.8535)
(0.4753,0.8980)
(0.4762,0.9214)
(0.4771,0.9149)
(0.4780,0.9199)
(0.4788,0.9188)
(0.4797,0.9247)
(0.4806,0.9286)
(0.4815,0.9342)
(0.4824,0.9514)
(0.4833,0.9312)
(0.4842,0.8810)
(0.4850,0.8704)
(0.4859,0.8969)
(0.4868,0.9226)
(0.4877,0.9404)
(0.4886,0.9433)
(0.4895,0.9306)
(0.4904,0.9351)
(0.4913,0.9278)
(0.4921,0.9280)
(0.4930,0.9278)
(0.4939,0.9073)
(0.4948,0.8944)
(0.4957,0.8853)
(0.4966,0.9020)
(0.4975,0.9110)
(0.4983,0.9101)
(0.4992,0.9083)
(0.5001,0.8894)
(0.5010,0.9172)
(0.5019,0.9262)
(0.5028,0.9051)
\PST@Solid(0.5028,0.9051)
(0.5037,0.8959)
(0.5046,0.8960)
(0.5054,0.8851)
(0.5063,0.8932)
(0.5072,0.8981)
(0.5081,0.8741)
(0.5090,0.8719)
(0.5099,0.8911)
(0.5108,0.9043)
(0.5116,0.8771)
(0.5125,0.8608)
(0.5134,0.8713)
(0.5143,0.8701)
(0.5152,0.8769)
(0.5161,0.9054)
(0.5170,0.9183)
(0.5179,0.9045)
(0.5187,0.8768)
(0.5196,0.8794)
(0.5205,0.8877)
(0.5214,0.8753)
(0.5223,0.8873)
(0.5232,0.8892)
(0.5241,0.8794)
(0.5249,0.8873)
(0.5258,0.9007)
(0.5267,0.9053)
(0.5276,0.8834)
(0.5285,0.8584)
(0.5294,0.8537)
(0.5303,0.8488)
(0.5312,0.8611)
(0.5320,0.8823)
(0.5329,0.8790)
(0.5338,0.8600)
(0.5347,0.8491)
(0.5356,0.8637)
(0.5365,0.8641)
(0.5374,0.8616)
(0.5382,0.8425)
(0.5391,0.8364)
(0.5400,0.8639)
(0.5409,0.8711)
(0.5418,0.8604)
(0.5427,0.8426)
(0.5436,0.8328)
(0.5445,0.8503)
(0.5453,0.8552)
(0.5462,0.8615)
(0.5471,0.8522)
(0.5480,0.8578)
(0.5489,0.8574)
(0.5498,0.8509)
(0.5507,0.8618)
(0.5515,0.8664)
(0.5524,0.8607)
(0.5533,0.8521)
(0.5542,0.8522)
(0.5551,0.8681)
(0.5560,0.8781)
(0.5569,0.8759)
(0.5578,0.8691)
(0.5586,0.8518)
(0.5595,0.8343)
(0.5604,0.8381)
(0.5613,0.8318)
(0.5622,0.8464)
(0.5631,0.8564)
(0.5640,0.8361)
(0.5648,0.8296)
(0.5657,0.8070)
(0.5666,0.8139)
(0.5675,0.8226)
(0.5684,0.8228)
(0.5693,0.8158)
(0.5702,0.7979)
(0.5711,0.8122)
(0.5719,0.8193)
(0.5728,0.8006)
(0.5737,0.8056)
(0.5746,0.8168)
(0.5755,0.7943)
(0.5764,0.7642)
(0.5773,0.7302)
(0.5781,0.6898)
(0.5790,0.6374)
(0.5799,0.5426)

\rput[r](0.8200,0.8010){$100\Delta$ -- Havege}
\PST@Dashed(0.8360,0.8010)
(0.9150,0.8010)

\PST@Dashed(0.1490,0.2100)
(0.1490,0.2100)
(0.1499,0.2194)
(0.1508,0.2219)
(0.1517,0.2373)
(0.1525,0.2259)
(0.1534,0.2153)
(0.1543,0.2169)
(0.1552,0.2153)
(0.1561,0.2179)
(0.1570,0.2169)
(0.1579,0.2153)
(0.1588,0.2103)
(0.1596,0.2128)
(0.1605,0.2169)
(0.1614,0.2103)
(0.1623,0.2194)
(0.1632,0.2128)
(0.1641,0.2103)
(0.1650,0.2103)
(0.1658,0.2103)
(0.1667,0.2128)
(0.1676,0.2153)
(0.1685,0.2179)
(0.1694,0.2179)
(0.1703,0.2204)
(0.1712,0.2219)
(0.1721,0.2103)
(0.1729,0.2179)
(0.1738,0.2169)
(0.1747,0.2153)
(0.1756,0.2244)
(0.1765,0.2128)
(0.1774,0.2153)
(0.1783,0.2103)
(0.1791,0.2128)
(0.1800,0.2259)
(0.1809,0.2153)
(0.1818,0.2219)
(0.1827,0.2128)
(0.1836,0.2103)
(0.1845,0.2153)
(0.1854,0.2128)
(0.1862,0.2179)
(0.1871,0.2153)
(0.1880,0.2219)
(0.1889,0.2128)
(0.1898,0.2103)
(0.1907,0.2103)
(0.1916,0.2103)
(0.1924,0.2219)
(0.1933,0.2270)
(0.1942,0.2103)
(0.1951,0.2103)
(0.1960,0.2219)
(0.1969,0.2153)
(0.1978,0.2103)
(0.1987,0.2194)
(0.1995,0.2153)
(0.2004,0.2244)
(0.2013,0.2277)
(0.2022,0.2153)
(0.2031,0.2128)
(0.2040,0.2169)
(0.2049,0.2153)
(0.2057,0.2169)
(0.2066,0.2244)
(0.2075,0.2169)
(0.2084,0.2103)
(0.2093,0.2128)
(0.2102,0.2103)
(0.2111,0.2128)
(0.2120,0.2103)
(0.2128,0.2128)
(0.2137,0.2219)
(0.2146,0.2270)
(0.2155,0.2259)
(0.2164,0.2310)
(0.2173,0.2128)
(0.2182,0.2179)
(0.2190,0.2229)
(0.2199,0.2103)
(0.2208,0.2128)
(0.2217,0.2128)
(0.2226,0.2270)
(0.2235,0.2270)
(0.2244,0.2179)
(0.2253,0.2103)
(0.2261,0.2153)
(0.2270,0.2259)
(0.2279,0.2219)
(0.2288,0.2194)
(0.2297,0.2310)
(0.2306,0.2194)
(0.2315,0.2285)
(0.2323,0.2547)
(0.2332,0.2659)
(0.2341,0.2219)
(0.2350,0.2244)
(0.2359,0.2153)
(0.2368,0.2153)
\PST@Dashed(0.2368,0.2153)
(0.2377,0.2244)
(0.2386,0.2219)
(0.2394,0.2219)
(0.2403,0.2229)
(0.2412,0.2270)
(0.2421,0.2244)
(0.2430,0.2128)
(0.2439,0.2128)
(0.2448,0.2128)
(0.2456,0.2401)
(0.2465,0.2608)
(0.2474,0.2535)
(0.2483,0.2477)
(0.2492,0.2295)
(0.2501,0.2360)
(0.2510,0.2179)
(0.2519,0.2270)
(0.2527,0.2282)
(0.2536,0.2194)
(0.2545,0.2335)
(0.2554,0.2371)
(0.2563,0.2477)
(0.2572,0.2330)
(0.2581,0.2217)
(0.2589,0.2320)
(0.2598,0.2570)
(0.2607,0.2153)
(0.2616,0.2194)
(0.2625,0.2244)
(0.2634,0.2328)
(0.2643,0.2277)
(0.2652,0.2285)
(0.2660,0.2328)
(0.2669,0.2259)
(0.2678,0.2320)
(0.2687,0.2310)
(0.2696,0.2153)
(0.2705,0.2270)
(0.2714,0.2489)
(0.2722,0.2530)
(0.2731,0.2499)
(0.2740,0.2636)
(0.2749,0.2520)
(0.2758,0.2257)
(0.2767,0.2171)
(0.2776,0.2643)
(0.2785,0.2401)
(0.2793,0.2179)
(0.2802,0.2333)
(0.2811,0.2270)
(0.2820,0.2525)
(0.2829,0.2444)
(0.2838,0.2530)
(0.2847,0.2128)
(0.2855,0.2128)
(0.2864,0.2128)
(0.2873,0.2166)
(0.2882,0.2128)
(0.2891,0.2179)
(0.2900,0.2153)
(0.2909,0.2179)
(0.2918,0.2194)
(0.2926,0.2179)
(0.2935,0.2242)
(0.2944,0.2179)
(0.2953,0.2179)
(0.2962,0.2153)
(0.2971,0.2219)
(0.2980,0.2434)
(0.2988,0.2244)
(0.2997,0.2219)
(0.3006,0.2128)
(0.3015,0.2131)
(0.3024,0.2179)
(0.3033,0.2194)
(0.3042,0.2242)
(0.3051,0.2103)
(0.3059,0.2179)
(0.3068,0.2386)
(0.3077,0.2330)
(0.3086,0.2383)
(0.3095,0.2244)
(0.3104,0.2393)
(0.3113,0.2179)
(0.3121,0.2179)
(0.3130,0.2169)
(0.3139,0.2179)
(0.3148,0.2270)
(0.3157,0.2320)
(0.3166,0.2219)
(0.3175,0.2285)
(0.3184,0.2153)
(0.3192,0.2194)
(0.3201,0.2128)
(0.3210,0.2179)
(0.3219,0.2244)
(0.3228,0.2128)
(0.3237,0.2219)
(0.3246,0.2219)
(0.3254,0.2302)
\PST@Dashed(0.3254,0.2302)
(0.3263,0.2128)
(0.3272,0.2103)
(0.3281,0.2103)
(0.3290,0.2128)
(0.3299,0.2194)
(0.3308,0.2103)
(0.3317,0.2103)
(0.3325,0.2219)
(0.3334,0.2194)
(0.3343,0.2153)
(0.3352,0.2179)
(0.3361,0.2103)
(0.3370,0.2103)
(0.3379,0.2128)
(0.3387,0.2128)
(0.3396,0.2128)
(0.3405,0.2103)
(0.3414,0.2128)
(0.3423,0.2103)
(0.3432,0.2128)
(0.3441,0.2103)
(0.3450,0.2153)
(0.3458,0.2169)
(0.3467,0.2103)
(0.3476,0.2229)
(0.3485,0.2219)
(0.3494,0.2153)
(0.3503,0.2128)
(0.3512,0.2234)
(0.3520,0.2128)
(0.3529,0.2153)
(0.3538,0.2103)
(0.3547,0.2103)
(0.3556,0.2128)
(0.3565,0.2103)
(0.3574,0.2103)
(0.3583,0.2103)
(0.3591,0.2103)
(0.3600,0.2103)
(0.3609,0.2153)
(0.3618,0.2128)
(0.3627,0.2103)
(0.3636,0.2103)
(0.3645,0.2285)
(0.3653,0.2179)
(0.3662,0.2128)
(0.3671,0.2103)
(0.3680,0.2103)
(0.3689,0.2103)
(0.3698,0.2103)
(0.3707,0.2103)
(0.3716,0.2103)
(0.3724,0.2128)
(0.3733,0.2194)
(0.3742,0.2128)
(0.3751,0.2103)
(0.3760,0.2103)
(0.3769,0.2128)
(0.3778,0.2128)
(0.3786,0.2128)
(0.3795,0.2153)
(0.3804,0.2128)
(0.3813,0.2103)
(0.3822,0.2103)
(0.3831,0.2103)
(0.3840,0.2103)
(0.3849,0.2103)
(0.3857,0.2103)
(0.3866,0.2103)
(0.3875,0.2103)
(0.3884,0.2153)
(0.3893,0.2103)
(0.3902,0.2103)
(0.3911,0.2103)
(0.3919,0.2103)
(0.3928,0.2103)
(0.3937,0.2103)
(0.3946,0.2128)
(0.3955,0.2103)
(0.3964,0.2103)
(0.3973,0.2128)
(0.3982,0.2103)
(0.3990,0.2103)
(0.3999,0.2103)
(0.4008,0.2128)
(0.4017,0.2103)
(0.4026,0.2103)
(0.4035,0.2128)
(0.4044,0.2103)
(0.4052,0.2103)
(0.4061,0.2103)
(0.4070,0.2153)
(0.4079,0.2103)
(0.4088,0.2128)
(0.4097,0.2219)
(0.4106,0.2153)
(0.4115,0.2128)
(0.4123,0.2103)
(0.4132,0.2103)
(0.4141,0.2103)
\PST@Dashed(0.4141,0.2103)
(0.4150,0.2128)
(0.4159,0.2103)
(0.4168,0.2103)
(0.4177,0.2103)
(0.4185,0.2103)
(0.4194,0.2103)
(0.4203,0.2103)
(0.4212,0.2103)
(0.4221,0.2103)
(0.4230,0.2128)
(0.4239,0.2103)
(0.4248,0.2103)
(0.4256,0.2128)
(0.4265,0.2103)
(0.4274,0.2103)
(0.4283,0.2103)
(0.4292,0.2103)
(0.4301,0.2103)
(0.4310,0.2103)
(0.4318,0.2103)
(0.4327,0.2103)
(0.4336,0.2103)
(0.4345,0.2103)
(0.4354,0.2103)
(0.4363,0.2103)
(0.4372,0.2103)
(0.4381,0.2128)
(0.4389,0.2153)
(0.4398,0.2128)
(0.4407,0.2103)
(0.4416,0.2103)
(0.4425,0.2103)
(0.4434,0.2103)
(0.4443,0.2128)
(0.4451,0.2128)
(0.4460,0.2103)
(0.4469,0.2103)
(0.4478,0.2103)
(0.4487,0.2100)
(0.4496,0.2103)
(0.4505,0.2105)
(0.4514,0.2123)
(0.4522,0.2123)
(0.4531,0.2090)
(0.4540,0.2128)
(0.4549,0.2126)
(0.4558,0.2083)
(0.4567,0.2083)
(0.4576,0.2126)
(0.4584,0.2189)
(0.4593,0.2151)
(0.4602,0.2113)
(0.4611,0.2128)
(0.4620,0.2108)
(0.4629,0.2126)
(0.4638,0.2118)
(0.4647,0.2093)
(0.4655,0.2093)
(0.4664,0.2126)
(0.4673,0.2103)
(0.4682,0.2103)
(0.4691,0.2100)
(0.4700,0.2103)
(0.4709,0.2105)
(0.4717,0.2103)
(0.4726,0.2098)
(0.4735,0.2088)
(0.4744,0.2083)
(0.4753,0.2090)
(0.4762,0.2128)
(0.4771,0.2131)
(0.4780,0.2113)
(0.4788,0.2073)
(0.4797,0.2075)
(0.4806,0.2098)
(0.4815,0.2098)
(0.4824,0.2133)
(0.4833,0.2123)
(0.4842,0.2128)
(0.4850,0.2103)
(0.4859,0.2131)
(0.4868,0.2110)
(0.4877,0.2105)
(0.4886,0.2103)
(0.4895,0.2103)
(0.4904,0.2115)
(0.4913,0.2115)
(0.4921,0.2131)
(0.4930,0.2085)
(0.4939,0.2133)
(0.4948,0.2118)
(0.4957,0.2118)
(0.4966,0.2098)
(0.4975,0.2110)
(0.4983,0.2105)
(0.4992,0.2070)
(0.5001,0.2080)
(0.5010,0.2093)
(0.5019,0.2136)
(0.5028,0.2095)
\PST@Dashed(0.5028,0.2095)
(0.5037,0.2118)
(0.5046,0.2110)
(0.5054,0.2131)
(0.5063,0.2110)
(0.5072,0.2093)
(0.5081,0.2103)
(0.5090,0.2100)
(0.5099,0.2075)
(0.5108,0.2105)
(0.5116,0.2105)
(0.5125,0.2128)
(0.5134,0.2098)
(0.5143,0.2133)
(0.5152,0.2121)
(0.5161,0.2098)
(0.5170,0.2115)
(0.5179,0.2103)
(0.5187,0.2133)
(0.5196,0.2103)
(0.5205,0.2083)
(0.5214,0.2121)
(0.5223,0.2136)
(0.5232,0.2105)
(0.5241,0.2103)
(0.5249,0.2108)
(0.5258,0.2105)
(0.5267,0.2126)
(0.5276,0.2110)
(0.5285,0.2126)
(0.5294,0.2133)
(0.5303,0.2133)
(0.5312,0.2078)
(0.5320,0.2136)
(0.5329,0.2133)
(0.5338,0.2121)
(0.5347,0.2078)
(0.5356,0.2123)
(0.5365,0.2073)
(0.5374,0.2100)
(0.5382,0.2088)
(0.5391,0.2118)
(0.5400,0.2118)
(0.5409,0.2075)
(0.5418,0.2080)
(0.5427,0.2088)
(0.5436,0.2123)
(0.5445,0.2126)
(0.5453,0.2115)
(0.5462,0.2118)
(0.5471,0.2080)
(0.5480,0.2070)
(0.5489,0.2100)
(0.5498,0.2108)
(0.5507,0.2098)
(0.5515,0.2098)
(0.5524,0.2083)
(0.5533,0.2088)
(0.5542,0.2126)
(0.5551,0.2128)
(0.5560,0.2123)
(0.5569,0.2118)
(0.5578,0.2100)
(0.5586,0.2083)
(0.5595,0.2115)
(0.5604,0.2095)
(0.5613,0.2085)
(0.5622,0.2085)
(0.5631,0.2083)
(0.5640,0.2110)
(0.5648,0.2110)
(0.5657,0.2126)
(0.5666,0.2085)
(0.5675,0.2078)
(0.5684,0.2095)
(0.5693,0.2131)
(0.5702,0.2090)
(0.5711,0.2073)
(0.5719,0.2083)
(0.5728,0.2118)
(0.5737,0.2083)
(0.5746,0.2083)
(0.5755,0.2121)
(0.5764,0.2105)
(0.5773,0.2103)
(0.5781,0.2100)
(0.5790,0.2103)
(0.5799,0.2103)

\rput[r](0.8200,0.9270){${\cal C}$ -- Normal RNG}
\PST@Dotted(0.8360,0.9270)
(0.9150,0.9270)

\PST@Dotted(0.1490,0.2104)
(0.1490,0.2104)
(0.1499,0.3298)
(0.1508,0.4160)
(0.1517,0.3140)
(0.1525,0.2323)
(0.1534,0.2284)
(0.1543,0.2305)
(0.1552,0.2484)
(0.1561,0.2596)
(0.1570,0.2555)
(0.1579,0.2430)
(0.1588,0.2474)
(0.1596,0.2561)
(0.1605,0.2441)
(0.1614,0.2500)
(0.1623,0.2567)
(0.1632,0.2567)
(0.1641,0.2546)
(0.1650,0.2366)
(0.1658,0.2338)
(0.1667,0.2455)
(0.1676,0.2537)
(0.1685,0.2478)
(0.1694,0.2377)
(0.1703,0.2366)
(0.1712,0.2483)
(0.1721,0.2688)
(0.1729,0.2772)
(0.1738,0.2842)
(0.1747,0.2751)
(0.1756,0.2634)
(0.1765,0.2530)
(0.1774,0.2493)
(0.1783,0.2673)
(0.1791,0.2876)
(0.1800,0.3322)
(0.1809,0.3303)
(0.1818,0.2973)
(0.1827,0.2931)
(0.1836,0.2803)
(0.1845,0.2740)
(0.1854,0.2764)
(0.1862,0.2790)
(0.1871,0.2955)
(0.1880,0.3026)
(0.1889,0.3104)
(0.1898,0.3139)
(0.1907,0.2827)
(0.1916,0.3069)
(0.1924,0.3137)
(0.1933,0.2768)
(0.1942,0.2990)
(0.1951,0.3334)
(0.1960,0.3246)
(0.1969,0.2956)
(0.1978,0.2698)
(0.1987,0.2855)
(0.1995,0.2921)
(0.2004,0.3026)
(0.2013,0.3083)
(0.2022,0.2951)
(0.2031,0.2859)
(0.2040,0.2715)
(0.2049,0.2874)
(0.2057,0.2884)
(0.2066,0.2580)
(0.2075,0.2654)
(0.2084,0.3111)
(0.2093,0.2862)
(0.2102,0.2462)
(0.2111,0.2699)
(0.2120,0.2829)
(0.2128,0.2580)
(0.2137,0.2398)
(0.2146,0.2383)
(0.2155,0.2322)
(0.2164,0.2329)
(0.2173,0.2639)
(0.2182,0.2670)
(0.2190,0.2752)
(0.2199,0.2863)
(0.2208,0.2583)
(0.2217,0.2552)
(0.2226,0.2560)
(0.2235,0.2738)
(0.2244,0.3108)
(0.2253,0.3489)
(0.2261,0.3603)
(0.2270,0.3173)
(0.2279,0.2943)
(0.2288,0.3235)
(0.2297,0.3006)
(0.2306,0.3892)
(0.2315,0.5690)
(0.2323,0.5580)
(0.2332,0.4562)
(0.2341,0.5155)
(0.2350,0.5332)
(0.2359,0.4129)
(0.2368,0.4394)
\PST@Dotted(0.2368,0.4394)
(0.2377,0.5172)
(0.2386,0.4781)
(0.2394,0.5352)
(0.2403,0.6060)
(0.2412,0.5595)
(0.2421,0.4366)
(0.2430,0.3185)
(0.2439,0.3036)
(0.2448,0.3533)
(0.2456,0.3960)
(0.2465,0.3523)
(0.2474,0.3424)
(0.2483,0.4832)
(0.2492,0.4971)
(0.2501,0.3879)
(0.2510,0.3500)
(0.2519,0.3473)
(0.2527,0.3177)
(0.2536,0.2738)
(0.2545,0.3134)
(0.2554,0.3229)
(0.2563,0.3377)
(0.2572,0.3466)
(0.2581,0.3509)
(0.2589,0.3607)
(0.2598,0.3165)
(0.2607,0.2962)
(0.2616,0.3339)
(0.2625,0.3597)
(0.2634,0.3104)
(0.2643,0.2709)
(0.2652,0.2843)
(0.2660,0.2929)
(0.2669,0.2880)
(0.2678,0.3249)
(0.2687,0.3484)
(0.2696,0.3287)
(0.2705,0.3628)
(0.2714,0.3729)
(0.2722,0.3459)
(0.2731,0.4027)
(0.2740,0.5033)
(0.2749,0.5239)
(0.2758,0.5064)
(0.2767,0.4538)
(0.2776,0.3626)
(0.2785,0.3106)
(0.2793,0.3482)
(0.2802,0.3847)
(0.2811,0.3515)
(0.2820,0.3638)
(0.2829,0.3766)
(0.2838,0.3295)
(0.2847,0.3031)
(0.2855,0.3081)
(0.2864,0.2837)
(0.2873,0.2935)
(0.2882,0.3352)
(0.2891,0.3670)
(0.2900,0.3611)
(0.2909,0.3382)
(0.2918,0.3257)
(0.2926,0.2947)
(0.2935,0.3053)
(0.2944,0.3232)
(0.2953,0.3270)
(0.2962,0.3200)
(0.2971,0.3114)
(0.2980,0.2964)
(0.2988,0.2993)
(0.2997,0.3014)
(0.3006,0.2765)
(0.3015,0.2605)
(0.3024,0.2419)
(0.3033,0.2394)
(0.3042,0.2673)
(0.3051,0.3043)
(0.3059,0.3104)
(0.3068,0.2714)
(0.3077,0.2389)
(0.3086,0.2380)
(0.3095,0.2526)
(0.3104,0.2773)
(0.3113,0.2813)
(0.3121,0.3374)
(0.3130,0.3554)
(0.3139,0.3135)
(0.3148,0.3025)
(0.3157,0.2888)
(0.3166,0.2444)
(0.3175,0.2371)
(0.3184,0.2465)
(0.3192,0.2451)
(0.3201,0.2507)
(0.3210,0.2494)
(0.3219,0.2574)
(0.3228,0.2747)
(0.3237,0.2676)
(0.3246,0.2557)
(0.3254,0.2716)
\PST@Dotted(0.3254,0.2716)
(0.3263,0.2745)
(0.3272,0.2662)
(0.3281,0.2656)
(0.3290,0.2683)
(0.3299,0.2879)
(0.3308,0.2940)
(0.3317,0.2922)
(0.3325,0.3093)
(0.3334,0.3201)
(0.3343,0.2891)
(0.3352,0.2548)
(0.3361,0.2746)
(0.3370,0.3105)
(0.3379,0.3284)
(0.3387,0.3083)
(0.3396,0.2855)
(0.3405,0.3069)
(0.3414,0.3007)
(0.3423,0.2898)
(0.3432,0.2988)
(0.3441,0.3214)
(0.3450,0.3147)
(0.3458,0.2556)
(0.3467,0.2158)
(0.3476,0.2198)
(0.3485,0.2253)
(0.3494,0.2421)
(0.3503,0.2618)
(0.3512,0.2730)
(0.3520,0.2894)
(0.3529,0.2791)
(0.3538,0.2905)
(0.3547,0.3094)
(0.3556,0.2651)
(0.3565,0.2449)
(0.3574,0.2534)
(0.3583,0.2706)
(0.3591,0.3063)
(0.3600,0.3192)
(0.3609,0.3142)
(0.3618,0.3112)
(0.3627,0.3061)
(0.3636,0.2969)
(0.3645,0.2898)
(0.3653,0.2855)
(0.3662,0.2867)
(0.3671,0.2998)
(0.3680,0.3283)
(0.3689,0.3660)
(0.3698,0.4094)
(0.3707,0.4509)
(0.3716,0.4863)
(0.3724,0.5121)
(0.3733,0.4803)
(0.3742,0.4236)
(0.3751,0.4168)
(0.3760,0.4076)
(0.3769,0.3786)
(0.3778,0.3839)
(0.3786,0.3744)
(0.3795,0.3880)
(0.3804,0.4065)
(0.3813,0.3881)
(0.3822,0.3440)
(0.3831,0.3269)
(0.3840,0.3111)
(0.3849,0.3104)
(0.3857,0.3217)
(0.3866,0.3264)
(0.3875,0.3466)
(0.3884,0.3476)
(0.3893,0.3370)
(0.3902,0.3441)
(0.3911,0.3480)
(0.3919,0.3396)
(0.3928,0.3303)
(0.3937,0.3205)
(0.3946,0.3195)
(0.3955,0.3000)
(0.3964,0.2785)
(0.3973,0.2921)
(0.3982,0.3159)
(0.3990,0.3137)
(0.3999,0.3160)
(0.4008,0.3027)
(0.4017,0.3043)
(0.4026,0.3007)
(0.4035,0.2905)
(0.4044,0.2967)
(0.4052,0.2954)
(0.4061,0.2927)
(0.4070,0.2973)
(0.4079,0.2956)
(0.4088,0.2878)
(0.4097,0.2869)
(0.4106,0.2950)
(0.4115,0.3068)
(0.4123,0.3088)
(0.4132,0.2975)
(0.4141,0.2886)
\PST@Dotted(0.4141,0.2886)
(0.4150,0.3013)
(0.4159,0.3066)
(0.4168,0.2991)
(0.4177,0.2950)
(0.4185,0.3057)
(0.4194,0.3008)
(0.4203,0.2881)
(0.4212,0.2812)
(0.4221,0.2795)
(0.4230,0.2762)
(0.4239,0.2795)
(0.4248,0.2831)
(0.4256,0.2828)
(0.4265,0.2750)
(0.4274,0.2556)
(0.4283,0.2623)
(0.4292,0.2640)
(0.4301,0.2604)
(0.4310,0.2596)
(0.4318,0.2585)
(0.4327,0.2673)
(0.4336,0.2717)
(0.4345,0.2731)
(0.4354,0.2689)
(0.4363,0.2632)
(0.4372,0.2652)
(0.4381,0.2695)
(0.4389,0.2708)
(0.4398,0.2685)
(0.4407,0.2681)
(0.4416,0.2704)
(0.4425,0.2699)
(0.4434,0.2661)
(0.4443,0.2678)
(0.4451,0.2721)
(0.4460,0.2697)
(0.4469,0.2647)
(0.4478,0.2634)
(0.4487,0.2600)
(0.4496,0.2611)
(0.4505,0.2666)
(0.4514,0.2657)
(0.4522,0.2632)
(0.4531,0.2698)
(0.4540,0.2729)
(0.4549,0.2674)
(0.4558,0.2721)
(0.4567,0.2687)
(0.4576,0.2672)
(0.4584,0.2746)
(0.4593,0.2742)
(0.4602,0.2761)
(0.4611,0.2792)
(0.4620,0.2830)
(0.4629,0.2824)
(0.4638,0.2788)
(0.4647,0.2688)
(0.4655,0.2688)
(0.4664,0.2823)
(0.4673,0.2827)
(0.4682,0.2696)
(0.4691,0.2659)
(0.4700,0.2622)
(0.4709,0.2639)
(0.4717,0.2664)
(0.4726,0.2619)
(0.4735,0.2559)
(0.4744,0.2578)
(0.4753,0.2638)
(0.4762,0.2641)
(0.4771,0.2609)
(0.4780,0.2596)
(0.4788,0.2592)
(0.4797,0.2545)
(0.4806,0.2518)
(0.4815,0.2536)
(0.4824,0.2651)
(0.4833,0.2667)
(0.4842,0.2607)
(0.4850,0.2601)
(0.4859,0.2697)
(0.4868,0.2776)
(0.4877,0.2770)
(0.4886,0.2764)
(0.4895,0.2776)
(0.4904,0.2854)
(0.4913,0.2934)
(0.4921,0.2938)
(0.4930,0.2945)
(0.4939,0.2908)
(0.4948,0.2802)
(0.4957,0.2737)
(0.4966,0.2747)
(0.4975,0.2778)
(0.4983,0.2772)
(0.4992,0.2814)
(0.5001,0.2877)
(0.5010,0.2869)
(0.5019,0.2811)
(0.5028,0.2859)
\PST@Dotted(0.5028,0.2859)
(0.5037,0.2913)
(0.5046,0.2948)
(0.5054,0.2931)
(0.5063,0.2896)
(0.5072,0.3004)
(0.5081,0.3062)
(0.5090,0.3056)
(0.5099,0.2950)
(0.5108,0.2996)
(0.5116,0.3049)
(0.5125,0.2986)
(0.5134,0.3024)
(0.5143,0.3033)
(0.5152,0.2952)
(0.5161,0.2921)
(0.5170,0.2919)
(0.5179,0.2872)
(0.5187,0.2772)
(0.5196,0.2800)
(0.5205,0.2800)
(0.5214,0.2777)
(0.5223,0.2758)
(0.5232,0.2748)
(0.5241,0.2745)
(0.5249,0.2792)
(0.5258,0.2842)
(0.5267,0.2799)
(0.5276,0.2770)
(0.5285,0.2797)
(0.5294,0.2784)
(0.5303,0.2792)
(0.5312,0.2775)
(0.5320,0.2791)
(0.5329,0.2857)
(0.5338,0.2860)
(0.5347,0.2776)
(0.5356,0.2742)
(0.5365,0.2765)
(0.5374,0.2797)
(0.5382,0.2844)
(0.5391,0.2783)
(0.5400,0.2861)
(0.5409,0.2878)
(0.5418,0.2945)
(0.5427,0.2967)
(0.5436,0.2957)
(0.5445,0.2887)
(0.5453,0.2822)
(0.5462,0.2832)
(0.5471,0.2874)
(0.5480,0.2851)
(0.5489,0.2821)
(0.5498,0.2813)
(0.5507,0.2859)
(0.5515,0.2810)
(0.5524,0.2760)
(0.5533,0.2768)
(0.5542,0.2762)
(0.5551,0.2810)
(0.5560,0.2860)
(0.5569,0.2816)
(0.5578,0.2786)
(0.5586,0.2797)
(0.5595,0.2849)
(0.5604,0.2929)
(0.5613,0.3013)
(0.5622,0.3013)
(0.5631,0.3079)
(0.5640,0.3069)
(0.5648,0.3013)
(0.5657,0.2964)
(0.5666,0.2864)
(0.5675,0.2804)
(0.5684,0.2836)
(0.5693,0.2861)
(0.5702,0.2911)
(0.5711,0.2902)
(0.5719,0.2883)
(0.5728,0.2839)
(0.5737,0.2806)
(0.5746,0.2738)
(0.5755,0.2785)
(0.5764,0.2977)
(0.5773,0.3619)
(0.5781,0.4340)
(0.5790,0.5013)
(0.5799,0.5246)
(0.5808,0.5199)
(0.5817,0.4999)
(0.5826,0.4933)
(0.5835,0.5073)
(0.5844,0.4999)
(0.5852,0.4975)
(0.5861,0.5130)
(0.5870,0.5158)
(0.5879,0.5027)
(0.5888,0.5078)
(0.5897,0.5228)
(0.5906,0.5142)
(0.5914,0.4924)
\PST@Dotted(0.5914,0.4924)
(0.5923,0.4846)
(0.5932,0.4908)
(0.5941,0.5086)
(0.5950,0.4942)
(0.5959,0.4830)
(0.5968,0.4954)
(0.5977,0.4872)
(0.5985,0.4838)
(0.5994,0.4885)
(0.6003,0.4971)
(0.6012,0.4965)
(0.6021,0.4804)
(0.6030,0.4820)
(0.6039,0.4971)
(0.6047,0.4908)
(0.6056,0.4888)
(0.6065,0.4936)
(0.6074,0.4853)
(0.6083,0.4841)
(0.6092,0.4911)
(0.6101,0.4820)
(0.6110,0.4819)
(0.6118,0.4819)
(0.6127,0.4944)
(0.6136,0.5089)
(0.6145,0.5194)
(0.6154,0.5234)
(0.6163,0.5237)
(0.6172,0.5411)
(0.6180,0.5408)
(0.6189,0.5408)
(0.6198,0.5552)
(0.6207,0.5546)
(0.6216,0.5669)
(0.6225,0.5584)
(0.6234,0.5396)
(0.6243,0.5285)
(0.6251,0.5337)
(0.6260,0.5263)
(0.6269,0.4959)
(0.6278,0.5021)
(0.6287,0.5272)
(0.6296,0.5244)
(0.6305,0.5261)
(0.6313,0.5389)
(0.6322,0.5498)
(0.6331,0.5682)
(0.6340,0.5644)
(0.6349,0.5377)
(0.6358,0.5415)
(0.6367,0.5372)
(0.6376,0.5140)
(0.6384,0.5203)
(0.6393,0.5276)
(0.6402,0.5234)
(0.6411,0.5153)
(0.6420,0.5175)
(0.6429,0.5404)
(0.6438,0.5468)
(0.6446,0.5352)
(0.6455,0.5401)
(0.6464,0.5292)
(0.6473,0.5178)
(0.6482,0.5198)
(0.6491,0.5178)
(0.6500,0.5179)
(0.6509,0.5213)
(0.6517,0.5161)
(0.6526,0.5099)
(0.6535,0.5175)
(0.6544,0.5296)
(0.6553,0.5419)
(0.6562,0.5319)
(0.6571,0.5327)
(0.6579,0.5332)
(0.6588,0.5168)
(0.6597,0.5124)
(0.6606,0.5082)
(0.6615,0.5081)
(0.6624,0.5095)
(0.6633,0.5123)
(0.6642,0.5147)
(0.6650,0.5319)
(0.6659,0.5362)
(0.6668,0.5300)
(0.6677,0.5257)
(0.6686,0.5142)
(0.6695,0.5138)
(0.6704,0.5206)
(0.6712,0.5287)
(0.6721,0.5350)
(0.6730,0.5370)
(0.6739,0.5622)
(0.6748,0.5665)
(0.6757,0.5429)
(0.6766,0.5220)
(0.6775,0.5225)
(0.6783,0.5133)
(0.6792,0.5197)
(0.6801,0.5156)
\PST@Dotted(0.6801,0.5156)
(0.6810,0.5030)
(0.6819,0.5282)
(0.6828,0.5317)
(0.6837,0.5268)
(0.6845,0.5197)
(0.6854,0.5184)
(0.6863,0.5168)
(0.6872,0.5192)
(0.6881,0.5320)
(0.6890,0.5332)
(0.6899,0.5386)
(0.6908,0.5409)
(0.6916,0.5247)
(0.6925,0.5166)
(0.6934,0.5283)
(0.6943,0.5363)
(0.6952,0.5218)
(0.6961,0.5136)
(0.6970,0.5069)
(0.6978,0.5179)
(0.6987,0.5163)
(0.6996,0.5174)
(0.7005,0.5483)
(0.7014,0.5660)
(0.7023,0.5776)
(0.7032,0.5836)
(0.7041,0.5821)
(0.7049,0.5844)
(0.7058,0.5929)
(0.7067,0.5997)
(0.7076,0.5852)
(0.7085,0.5842)
(0.7094,0.5945)
(0.7103,0.6050)
(0.7111,0.6145)
(0.7120,0.6002)
(0.7129,0.6140)
(0.7138,0.6081)
(0.7147,0.5900)
(0.7156,0.5827)
(0.7165,0.5717)
(0.7174,0.5598)
(0.7182,0.5475)
(0.7191,0.5544)
(0.7200,0.5517)
(0.7209,0.5581)
(0.7218,0.5504)
(0.7227,0.5431)
(0.7236,0.5499)
(0.7244,0.5658)
(0.7253,0.5515)
(0.7262,0.5476)
(0.7271,0.5452)
(0.7280,0.5532)
(0.7289,0.5559)
(0.7298,0.5503)
(0.7307,0.5597)
(0.7315,0.5672)
(0.7324,0.5511)
(0.7333,0.5298)
(0.7342,0.5429)
(0.7351,0.5732)
(0.7360,0.5812)
(0.7369,0.5863)
(0.7377,0.5605)
(0.7386,0.5525)
(0.7395,0.5537)
(0.7404,0.5321)
(0.7413,0.5150)
(0.7422,0.5165)
(0.7431,0.5033)
(0.7440,0.4903)
(0.7448,0.5047)
(0.7457,0.5193)
(0.7466,0.5165)
(0.7475,0.5077)
(0.7484,0.5116)
(0.7493,0.5202)
(0.7502,0.5204)
(0.7510,0.5360)
(0.7519,0.5445)
(0.7528,0.5444)
(0.7537,0.5598)
(0.7546,0.5665)
(0.7555,0.5594)
(0.7564,0.5614)
(0.7573,0.5717)
(0.7581,0.5723)
(0.7590,0.5651)
(0.7599,0.5365)
(0.7608,0.5103)
(0.7617,0.5147)
(0.7626,0.5080)
(0.7635,0.5021)
(0.7643,0.4956)
(0.7652,0.5104)
(0.7661,0.5018)
(0.7670,0.4994)
(0.7679,0.5227)
(0.7688,0.5275)
\PST@Dotted(0.7688,0.5275)
(0.7697,0.5329)
(0.7706,0.5257)
(0.7714,0.5164)
(0.7723,0.5248)
(0.7732,0.5376)
(0.7741,0.5443)
(0.7750,0.5503)
(0.7759,0.5530)
(0.7768,0.5616)
(0.7776,0.5676)
(0.7785,0.5477)
(0.7794,0.5323)
(0.7803,0.5163)
(0.7812,0.5021)
(0.7821,0.5134)
(0.7830,0.5123)
(0.7839,0.5301)
(0.7847,0.5233)
(0.7856,0.5208)
(0.7865,0.5210)
(0.7874,0.5037)
(0.7883,0.5048)
(0.7892,0.4879)
(0.7901,0.4730)
(0.7909,0.4868)
(0.7918,0.5030)
(0.7927,0.5019)
(0.7936,0.4915)
(0.7945,0.4944)
(0.7954,0.4975)
(0.7963,0.5130)
(0.7972,0.5269)
(0.7980,0.5335)
(0.7989,0.5308)
(0.7998,0.5308)
(0.8007,0.5281)
(0.8016,0.4854)
(0.8025,0.4657)
(0.8034,0.4614)
(0.8042,0.4666)
(0.8051,0.4679)
(0.8060,0.4810)
(0.8069,0.4884)
(0.8078,0.4822)
(0.8087,0.4862)
(0.8096,0.4718)
(0.8105,0.4690)
(0.8113,0.4696)
(0.8122,0.4683)
(0.8131,0.4863)
(0.8140,0.4891)
(0.8149,0.4992)
(0.8158,0.5093)
(0.8167,0.5018)
(0.8175,0.4854)
(0.8184,0.4866)
(0.8193,0.5012)
(0.8202,0.4989)
(0.8211,0.4909)
(0.8220,0.4797)
(0.8229,0.4677)
(0.8238,0.4699)
(0.8246,0.4909)
(0.8255,0.4928)
(0.8264,0.4612)
(0.8273,0.4476)
(0.8282,0.4697)
(0.8291,0.4869)
(0.8300,0.4823)
(0.8308,0.4766)
(0.8317,0.4775)
(0.8326,0.4774)
(0.8335,0.4740)
(0.8344,0.4676)
(0.8353,0.4590)
(0.8362,0.4623)
(0.8371,0.4574)
(0.8379,0.4565)
(0.8388,0.4626)
(0.8397,0.4535)
(0.8406,0.4572)
(0.8415,0.4444)
(0.8424,0.4425)
(0.8433,0.4494)
(0.8441,0.4680)
(0.8450,0.4710)
(0.8459,0.4550)
(0.8468,0.4414)
(0.8477,0.4378)
(0.8486,0.4403)
(0.8495,0.4438)
(0.8504,0.4298)
(0.8512,0.4238)
(0.8521,0.4368)
(0.8530,0.4494)
(0.8539,0.4352)
(0.8548,0.4275)
(0.8557,0.4427)
(0.8566,0.4626)
(0.8574,0.4571)
\PST@Dotted(0.8574,0.4571)
(0.8583,0.4335)
(0.8592,0.4316)
(0.8601,0.4269)
(0.8610,0.4342)
(0.8619,0.4394)
(0.8628,0.4289)
(0.8637,0.4061)
(0.8645,0.3684)
(0.8654,0.3510)
(0.8663,0.3190)
(0.8672,0.2625)

\rput[r](0.8200,0.8850){$100\Delta$ -- Normal RNG}
\PST@LongDash(0.8360,0.8850)
(0.9150,0.8850)

\PST@LongDash(0.1490,0.2103)
(0.1490,0.2103)
(0.1499,0.2270)
(0.1508,0.2270)
(0.1517,0.2353)
(0.1525,0.2176)
(0.1534,0.2103)
(0.1543,0.2153)
(0.1552,0.2128)
(0.1561,0.2128)
(0.1570,0.2515)
(0.1579,0.2282)
(0.1588,0.2557)
(0.1596,0.2419)
(0.1605,0.2515)
(0.1614,0.2219)
(0.1623,0.2229)
(0.1632,0.2259)
(0.1641,0.2277)
(0.1650,0.2194)
(0.1658,0.2252)
(0.1667,0.2219)
(0.1676,0.2219)
(0.1685,0.2219)
(0.1694,0.2151)
(0.1703,0.2267)
(0.1712,0.2345)
(0.1721,0.2295)
(0.1729,0.2310)
(0.1738,0.2179)
(0.1747,0.2320)
(0.1756,0.2502)
(0.1765,0.2219)
(0.1774,0.2128)
(0.1783,0.2153)
(0.1791,0.2335)
(0.1800,0.2194)
(0.1809,0.2128)
(0.1818,0.2179)
(0.1827,0.2153)
(0.1836,0.2128)
(0.1845,0.2449)
(0.1854,0.2403)
(0.1862,0.2153)
(0.1871,0.2103)
(0.1880,0.2244)
(0.1889,0.2244)
(0.1898,0.2295)
(0.1907,0.2376)
(0.1916,0.2179)
(0.1924,0.2244)
(0.1933,0.2270)
(0.1942,0.2153)
(0.1951,0.2128)
(0.1960,0.2103)
(0.1969,0.2194)
(0.1978,0.2320)
(0.1987,0.2153)
(0.1995,0.2128)
(0.2004,0.2153)
(0.2013,0.2153)
(0.2022,0.2179)
(0.2031,0.2270)
(0.2040,0.2259)
(0.2049,0.2234)
(0.2057,0.2277)
(0.2066,0.2360)
(0.2075,0.2244)
(0.2084,0.2310)
(0.2093,0.2345)
(0.2102,0.2509)
(0.2111,0.2277)
(0.2120,0.2328)
(0.2128,0.2603)
(0.2137,0.2701)
(0.2146,0.2545)
(0.2155,0.2444)
(0.2164,0.2724)
(0.2173,0.2515)
(0.2182,0.2368)
(0.2190,0.2204)
(0.2199,0.2244)
(0.2208,0.2411)
(0.2217,0.2335)
(0.2226,0.2419)
(0.2235,0.2459)
(0.2244,0.2462)
(0.2253,0.2285)
(0.2261,0.2229)
(0.2270,0.2295)
(0.2279,0.2270)
(0.2288,0.2254)
(0.2297,0.2179)
(0.2306,0.2270)
(0.2315,0.2128)
(0.2323,0.2128)
(0.2332,0.2128)
(0.2341,0.2151)
(0.2350,0.2103)
(0.2359,0.2153)
(0.2368,0.2128)
\PST@LongDash(0.2368,0.2128)
(0.2377,0.2103)
(0.2386,0.2153)
(0.2394,0.2171)
(0.2403,0.2105)
(0.2412,0.2171)
(0.2421,0.2128)
(0.2430,0.2128)
(0.2439,0.2204)
(0.2448,0.2128)
(0.2456,0.2128)
(0.2465,0.2128)
(0.2474,0.2128)
(0.2483,0.2103)
(0.2492,0.2103)
(0.2501,0.2153)
(0.2510,0.2153)
(0.2519,0.2153)
(0.2527,0.2179)
(0.2536,0.2254)
(0.2545,0.2684)
(0.2554,0.2335)
(0.2563,0.2103)
(0.2572,0.2128)
(0.2581,0.2219)
(0.2589,0.2244)
(0.2598,0.2310)
(0.2607,0.2244)
(0.2616,0.2343)
(0.2625,0.2153)
(0.2634,0.2408)
(0.2643,0.2302)
(0.2652,0.2525)
(0.2660,0.2419)
(0.2669,0.2295)
(0.2678,0.2219)
(0.2687,0.2285)
(0.2696,0.2194)
(0.2705,0.2128)
(0.2714,0.2103)
(0.2722,0.2302)
(0.2731,0.2169)
(0.2740,0.2153)
(0.2749,0.2166)
(0.2758,0.2128)
(0.2767,0.2277)
(0.2776,0.2204)
(0.2785,0.2153)
(0.2793,0.2219)
(0.2802,0.2153)
(0.2811,0.2103)
(0.2820,0.2169)
(0.2829,0.2219)
(0.2838,0.2408)
(0.2847,0.2194)
(0.2855,0.2244)
(0.2864,0.2595)
(0.2873,0.2414)
(0.2882,0.2295)
(0.2891,0.2328)
(0.2900,0.2204)
(0.2909,0.2128)
(0.2918,0.2219)
(0.2926,0.2153)
(0.2935,0.2219)
(0.2944,0.2229)
(0.2953,0.2194)
(0.2962,0.2285)
(0.2971,0.2219)
(0.2980,0.2153)
(0.2988,0.2128)
(0.2997,0.2128)
(0.3006,0.2128)
(0.3015,0.2234)
(0.3024,0.2169)
(0.3033,0.2153)
(0.3042,0.2153)
(0.3051,0.2128)
(0.3059,0.2128)
(0.3068,0.2153)
(0.3077,0.2401)
(0.3086,0.2285)
(0.3095,0.2517)
(0.3104,0.2194)
(0.3113,0.2204)
(0.3121,0.2128)
(0.3130,0.2128)
(0.3139,0.2179)
(0.3148,0.2153)
(0.3157,0.2169)
(0.3166,0.2179)
(0.3175,0.2194)
(0.3184,0.2194)
(0.3192,0.2128)
(0.3201,0.2300)
(0.3210,0.2295)
(0.3219,0.2270)
(0.3228,0.2244)
(0.3237,0.2244)
(0.3246,0.2302)
(0.3254,0.2219)
\PST@LongDash(0.3254,0.2219)
(0.3263,0.2219)
(0.3272,0.2302)
(0.3281,0.2179)
(0.3290,0.2194)
(0.3299,0.2573)
(0.3308,0.2393)
(0.3317,0.2295)
(0.3325,0.2328)
(0.3334,0.2153)
(0.3343,0.2368)
(0.3352,0.2464)
(0.3361,0.2320)
(0.3370,0.2419)
(0.3379,0.2464)
(0.3387,0.2343)
(0.3396,0.2439)
(0.3405,0.2328)
(0.3414,0.2419)
(0.3423,0.2244)
(0.3432,0.2469)
(0.3441,0.2353)
(0.3450,0.2555)
(0.3458,0.2484)
(0.3467,0.2275)
(0.3476,0.2335)
(0.3485,0.2515)
(0.3494,0.2244)
(0.3503,0.2439)
(0.3512,0.2747)
(0.3520,0.2219)
(0.3529,0.2244)
(0.3538,0.2259)
(0.3547,0.2328)
(0.3556,0.2704)
(0.3565,0.2631)
(0.3574,0.2525)
(0.3583,0.2484)
(0.3591,0.2295)
(0.3600,0.2194)
(0.3609,0.2474)
(0.3618,0.2244)
(0.3627,0.2469)
(0.3636,0.2368)
(0.3645,0.2575)
(0.3653,0.2595)
(0.3662,0.3149)
(0.3671,0.2393)
(0.3680,0.2489)
(0.3689,0.2302)
(0.3698,0.2128)
(0.3707,0.2128)
(0.3716,0.2103)
(0.3724,0.2103)
(0.3733,0.2103)
(0.3742,0.2103)
(0.3751,0.2128)
(0.3760,0.2217)
(0.3769,0.2244)
(0.3778,0.2128)
(0.3786,0.2128)
(0.3795,0.2103)
(0.3804,0.2128)
(0.3813,0.2153)
(0.3822,0.2194)
(0.3831,0.2459)
(0.3840,0.2580)
(0.3849,0.2707)
(0.3857,0.2323)
(0.3866,0.2285)
(0.3875,0.2128)
(0.3884,0.2204)
(0.3893,0.2103)
(0.3902,0.2295)
(0.3911,0.2103)
(0.3919,0.2219)
(0.3928,0.2194)
(0.3937,0.2194)
(0.3946,0.2219)
(0.3955,0.2323)
(0.3964,0.2762)
(0.3973,0.2419)
(0.3982,0.2128)
(0.3990,0.2285)
(0.3999,0.2270)
(0.4008,0.2128)
(0.4017,0.2194)
(0.4026,0.2270)
(0.4035,0.2525)
(0.4044,0.2464)
(0.4052,0.2489)
(0.4061,0.2254)
(0.4070,0.2229)
(0.4079,0.2295)
(0.4088,0.2378)
(0.4097,0.2509)
(0.4106,0.2368)
(0.4115,0.2343)
(0.4123,0.2219)
(0.4132,0.2194)
(0.4141,0.2169)
\PST@LongDash(0.4141,0.2169)
(0.4150,0.2219)
(0.4159,0.2285)
(0.4168,0.2295)
(0.4177,0.2244)
(0.4185,0.2153)
(0.4194,0.2153)
(0.4203,0.2153)
(0.4212,0.2376)
(0.4221,0.2459)
(0.4230,0.2219)
(0.4239,0.2295)
(0.4248,0.2128)
(0.4256,0.2244)
(0.4265,0.2360)
(0.4274,0.2643)
(0.4283,0.2795)
(0.4292,0.2444)
(0.4301,0.2605)
(0.4310,0.2469)
(0.4318,0.2810)
(0.4327,0.2434)
(0.4336,0.2343)
(0.4345,0.2368)
(0.4354,0.2368)
(0.4363,0.2754)
(0.4372,0.2679)
(0.4381,0.2593)
(0.4389,0.2393)
(0.4398,0.2328)
(0.4407,0.2408)
(0.4416,0.2386)
(0.4425,0.2414)
(0.4434,0.2451)
(0.4443,0.2449)
(0.4451,0.2464)
(0.4460,0.2638)
(0.4469,0.2401)
(0.4478,0.2504)
(0.4487,0.2560)
(0.4496,0.2560)
(0.4505,0.2535)
(0.4514,0.2484)
(0.4522,0.2555)
(0.4531,0.2653)
(0.4540,0.2219)
(0.4549,0.2550)
(0.4558,0.2509)
(0.4567,0.2393)
(0.4576,0.2393)
(0.4584,0.2464)
(0.4593,0.2169)
(0.4602,0.2310)
(0.4611,0.2335)
(0.4620,0.2295)
(0.4629,0.2277)
(0.4638,0.2484)
(0.4647,0.2477)
(0.4655,0.2580)
(0.4664,0.2219)
(0.4673,0.2259)
(0.4682,0.2128)
(0.4691,0.2270)
(0.4700,0.2515)
(0.4709,0.2850)
(0.4717,0.2535)
(0.4726,0.2310)
(0.4735,0.2484)
(0.4744,0.2671)
(0.4753,0.2585)
(0.4762,0.2328)
(0.4771,0.2515)
(0.4780,0.2459)
(0.4788,0.2621)
(0.4797,0.2550)
(0.4806,0.2671)
(0.4815,0.2419)
(0.4824,0.2169)
(0.4833,0.2259)
(0.4842,0.2204)
(0.4850,0.2419)
(0.4859,0.2244)
(0.4868,0.2153)
(0.4877,0.2169)
(0.4886,0.2285)
(0.4895,0.2153)
(0.4904,0.2153)
(0.4913,0.2128)
(0.4921,0.2128)
(0.4930,0.2219)
(0.4939,0.2204)
(0.4948,0.2270)
(0.4957,0.2219)
(0.4966,0.2194)
(0.4975,0.2204)
(0.4983,0.2219)
(0.4992,0.2153)
(0.5001,0.2103)
(0.5010,0.2310)
(0.5019,0.2128)
(0.5028,0.2179)
\PST@LongDash(0.5028,0.2179)
(0.5037,0.2244)
(0.5046,0.2179)
(0.5054,0.2153)
(0.5063,0.2153)
(0.5072,0.2194)
(0.5081,0.2219)
(0.5090,0.2169)
(0.5099,0.2153)
(0.5108,0.2153)
(0.5116,0.2270)
(0.5125,0.2204)
(0.5134,0.2128)
(0.5143,0.2103)
(0.5152,0.2128)
(0.5161,0.2103)
(0.5170,0.2103)
(0.5179,0.2194)
(0.5187,0.2153)
(0.5196,0.2325)
(0.5205,0.2219)
(0.5214,0.2204)
(0.5223,0.2320)
(0.5232,0.2194)
(0.5241,0.2320)
(0.5249,0.2169)
(0.5258,0.2153)
(0.5267,0.2320)
(0.5276,0.2353)
(0.5285,0.2285)
(0.5294,0.2285)
(0.5303,0.2244)
(0.5312,0.2285)
(0.5320,0.2153)
(0.5329,0.2153)
(0.5338,0.2419)
(0.5347,0.2310)
(0.5356,0.2153)
(0.5365,0.2285)
(0.5374,0.2179)
(0.5382,0.2219)
(0.5391,0.2128)
(0.5400,0.2219)
(0.5409,0.2234)
(0.5418,0.2194)
(0.5427,0.2128)
(0.5436,0.2234)
(0.5445,0.2179)
(0.5453,0.2219)
(0.5462,0.2219)
(0.5471,0.2244)
(0.5480,0.2153)
(0.5489,0.2103)
(0.5498,0.2179)
(0.5507,0.2103)
(0.5515,0.2169)
(0.5524,0.2244)
(0.5533,0.2368)
(0.5542,0.2295)
(0.5551,0.2153)
(0.5560,0.2128)
(0.5569,0.2285)
(0.5578,0.2295)
(0.5586,0.2219)
(0.5595,0.2153)
(0.5604,0.2194)
(0.5613,0.2128)
(0.5622,0.2153)
(0.5631,0.2179)
(0.5640,0.2194)
(0.5648,0.2153)
(0.5657,0.2434)
(0.5666,0.2244)
(0.5675,0.2128)
(0.5684,0.2300)
(0.5693,0.2128)
(0.5702,0.2219)
(0.5711,0.2194)
(0.5719,0.2323)
(0.5728,0.2103)
(0.5737,0.2179)
(0.5746,0.2204)
(0.5755,0.2179)
(0.5764,0.2128)
(0.5773,0.2153)
(0.5781,0.2194)
(0.5790,0.2126)
(0.5799,0.2128)
(0.5808,0.2105)
(0.5817,0.2128)
(0.5826,0.2103)
(0.5835,0.2153)
(0.5844,0.2103)
(0.5852,0.2103)
(0.5861,0.2128)
(0.5870,0.2103)
(0.5879,0.2128)
(0.5888,0.2103)
(0.5897,0.2103)
(0.5906,0.2103)
(0.5914,0.2128)
\PST@LongDash(0.5914,0.2128)
(0.5923,0.2103)
(0.5932,0.2103)
(0.5941,0.2103)
(0.5950,0.2103)
(0.5959,0.2100)
(0.5968,0.2103)
(0.5977,0.2103)
(0.5985,0.2128)
(0.5994,0.2103)
(0.6003,0.2166)
(0.6012,0.2103)
(0.6021,0.2103)
(0.6030,0.2103)
(0.6039,0.2103)
(0.6047,0.2100)
(0.6056,0.2103)
(0.6065,0.2103)
(0.6074,0.2128)
(0.6083,0.2100)
(0.6092,0.2169)
(0.6101,0.2153)
(0.6110,0.2128)
(0.6118,0.2128)
(0.6127,0.2103)
(0.6136,0.2126)
(0.6145,0.2103)
(0.6154,0.2103)
(0.6163,0.2128)
(0.6172,0.2128)
(0.6180,0.2105)
(0.6189,0.2100)
(0.6198,0.2100)
(0.6207,0.2105)
(0.6216,0.2105)
(0.6225,0.2103)
(0.6234,0.2103)
(0.6243,0.2103)
(0.6251,0.2103)
(0.6260,0.2128)
(0.6269,0.2128)
(0.6278,0.2103)
(0.6287,0.2103)
(0.6296,0.2128)
(0.6305,0.2103)
(0.6313,0.2103)
(0.6322,0.2100)
(0.6331,0.2103)
(0.6340,0.2103)
(0.6349,0.2194)
(0.6358,0.2103)
(0.6367,0.2103)
(0.6376,0.2103)
(0.6384,0.2105)
(0.6393,0.2103)
(0.6402,0.2128)
(0.6411,0.2128)
(0.6420,0.2103)
(0.6429,0.2105)
(0.6438,0.2105)
(0.6446,0.2103)
(0.6455,0.2105)
(0.6464,0.2103)
(0.6473,0.2103)
(0.6482,0.2126)
(0.6491,0.2128)
(0.6500,0.2103)
(0.6509,0.2103)
(0.6517,0.2103)
(0.6526,0.2103)
(0.6535,0.2105)
(0.6544,0.2100)
(0.6553,0.2128)
(0.6562,0.2100)
(0.6571,0.2103)
(0.6579,0.2103)
(0.6588,0.2105)
(0.6597,0.2103)
(0.6606,0.2103)
(0.6615,0.2103)
(0.6624,0.2103)
(0.6633,0.2103)
(0.6642,0.2103)
(0.6650,0.2103)
(0.6659,0.2103)
(0.6668,0.2105)
(0.6677,0.2103)
(0.6686,0.2103)
(0.6695,0.2103)
(0.6704,0.2103)
(0.6712,0.2103)
(0.6721,0.2128)
(0.6730,0.2128)
(0.6739,0.2100)
(0.6748,0.2103)
(0.6757,0.2128)
(0.6766,0.2128)
(0.6775,0.2103)
(0.6783,0.2103)
(0.6792,0.2103)
(0.6801,0.2103)
\PST@LongDash(0.6801,0.2103)
(0.6810,0.2103)
(0.6819,0.2169)
(0.6828,0.2103)
(0.6837,0.2103)
(0.6845,0.2105)
(0.6854,0.2105)
(0.6863,0.2103)
(0.6872,0.2103)
(0.6881,0.2131)
(0.6890,0.2103)
(0.6899,0.2100)
(0.6908,0.2103)
(0.6916,0.2103)
(0.6925,0.2103)
(0.6934,0.2128)
(0.6943,0.2128)
(0.6952,0.2131)
(0.6961,0.2103)
(0.6970,0.2103)
(0.6978,0.2103)
(0.6987,0.2103)
(0.6996,0.2131)
(0.7005,0.2100)
(0.7014,0.2103)
(0.7023,0.2105)
(0.7032,0.2100)
(0.7041,0.2103)
(0.7049,0.2103)
(0.7058,0.2103)
(0.7067,0.2105)
(0.7076,0.2103)
(0.7085,0.2103)
(0.7094,0.2103)
(0.7103,0.2128)
(0.7111,0.2103)
(0.7120,0.2103)
(0.7129,0.2128)
(0.7138,0.2126)
(0.7147,0.2103)
(0.7156,0.2103)
(0.7165,0.2105)
(0.7174,0.2131)
(0.7182,0.2103)
(0.7191,0.2105)
(0.7200,0.2103)
(0.7209,0.2103)
(0.7218,0.2103)
(0.7227,0.2103)
(0.7236,0.2100)
(0.7244,0.2103)
(0.7253,0.2103)
(0.7262,0.2103)
(0.7271,0.2103)
(0.7280,0.2103)
(0.7289,0.2103)
(0.7298,0.2103)
(0.7307,0.2103)
(0.7315,0.2128)
(0.7324,0.2166)
(0.7333,0.2103)
(0.7342,0.2100)
(0.7351,0.2105)
(0.7360,0.2100)
(0.7369,0.2105)
(0.7377,0.2103)
(0.7386,0.2103)
(0.7395,0.2103)
(0.7404,0.2103)
(0.7413,0.2128)
(0.7422,0.2103)
(0.7431,0.2103)
(0.7440,0.2103)
(0.7448,0.2103)
(0.7457,0.2153)
(0.7466,0.2128)
(0.7475,0.2103)
(0.7484,0.2103)
(0.7493,0.2100)
(0.7502,0.2103)
(0.7510,0.2103)
(0.7519,0.2105)
(0.7528,0.2100)
(0.7537,0.2126)
(0.7546,0.2103)
(0.7555,0.2128)
(0.7564,0.2103)
(0.7573,0.2103)
(0.7581,0.2103)
(0.7590,0.2103)
(0.7599,0.2103)
(0.7608,0.2103)
(0.7617,0.2103)
(0.7626,0.2103)
(0.7635,0.2100)
(0.7643,0.2128)
(0.7652,0.2103)
(0.7661,0.2103)
(0.7670,0.2128)
(0.7679,0.2131)
(0.7688,0.2105)
\PST@LongDash(0.7688,0.2105)
(0.7697,0.2103)
(0.7706,0.2103)
(0.7714,0.2103)
(0.7723,0.2131)
(0.7732,0.2103)
(0.7741,0.2103)
(0.7750,0.2103)
(0.7759,0.2103)
(0.7768,0.2103)
(0.7776,0.2128)
(0.7785,0.2126)
(0.7794,0.2103)
(0.7803,0.2103)
(0.7812,0.2103)
(0.7821,0.2103)
(0.7830,0.2103)
(0.7839,0.2103)
(0.7847,0.2128)
(0.7856,0.2103)
(0.7865,0.2153)
(0.7874,0.2103)
(0.7883,0.2103)
(0.7892,0.2103)
(0.7901,0.2103)
(0.7909,0.2103)
(0.7918,0.2103)
(0.7927,0.2103)
(0.7936,0.2103)
(0.7945,0.2128)
(0.7954,0.2103)
(0.7963,0.2103)
(0.7972,0.2103)
(0.7980,0.2103)
(0.7989,0.2103)
(0.7998,0.2103)
(0.8007,0.2103)
(0.8016,0.2103)
(0.8025,0.2103)
(0.8034,0.2103)
(0.8042,0.2103)
(0.8051,0.2103)
(0.8060,0.2103)
(0.8069,0.2103)
(0.8078,0.2103)
(0.8087,0.2103)
(0.8096,0.2103)
(0.8105,0.2103)
(0.8113,0.2103)
(0.8122,0.2103)
(0.8131,0.2103)
(0.8140,0.2103)
(0.8149,0.2103)
(0.8158,0.2103)
(0.8167,0.2103)
(0.8175,0.2103)
(0.8184,0.2103)
(0.8193,0.2103)
(0.8202,0.2103)
(0.8211,0.2131)
(0.8220,0.2131)
(0.8229,0.2128)
(0.8238,0.2128)
(0.8246,0.2128)
(0.8255,0.2103)
(0.8264,0.2103)
(0.8273,0.2103)
(0.8282,0.2103)
(0.8291,0.2103)
(0.8300,0.2103)
(0.8308,0.2103)
(0.8317,0.2100)
(0.8326,0.2103)
(0.8335,0.2103)
(0.8344,0.2103)
(0.8353,0.2103)
(0.8362,0.2103)
(0.8371,0.2103)
(0.8379,0.2103)
(0.8388,0.2103)
(0.8397,0.2103)
(0.8406,0.2103)
(0.8415,0.2103)
(0.8424,0.2103)
(0.8433,0.2103)
(0.8441,0.2128)
(0.8450,0.2103)
(0.8459,0.2103)
(0.8468,0.2103)
(0.8477,0.2103)
(0.8486,0.2103)
(0.8495,0.2103)
(0.8504,0.2103)
(0.8512,0.2103)
(0.8521,0.2103)
(0.8530,0.2103)
(0.8539,0.2103)
(0.8548,0.2128)
(0.8557,0.2103)
(0.8566,0.2103)
(0.8574,0.2103)
\PST@LongDash(0.8574,0.2103)
(0.8583,0.2103)
(0.8592,0.2103)
(0.8601,0.2103)
(0.8610,0.2103)
(0.8619,0.2128)
(0.8628,0.2103)
(0.8637,0.2103)
(0.8645,0.2103)
(0.8654,0.2103)
(0.8663,0.2103)
(0.8672,0.2153)

\PST@Border(0.1490,0.9680)
(0.1490,0.0840)
(0.9470,0.0840)
(0.9470,0.9680)
(0.1490,0.9680)

\catcode`@=12
\fi
\endpspicture

%% file: ecolab.complexity.tex
\ifx\PSTloaded\undefined
\def\PSTloaded{t}
\psset{arrowsize=.01 3.2 1.4 .3}
\psset{dotsize=.01}
\catcode`@=11

\newpsobject{PST@Border}{psline}{linewidth=.0015,linestyle=solid}
\newpsobject{PST@Axes}{psline}{linewidth=.0015,linestyle=dotted,dotsep=.004}
\newpsobject{PST@Solid}{psline}{linewidth=.0015,linestyle=solid}
\newpsobject{PST@Dashed}{psline}{linewidth=.0015,linestyle=dashed,dash=.01 .01}
\newpsobject{PST@Dotted}{psline}{linewidth=.0025,linestyle=dotted,dotsep=.008}
\newpsobject{PST@LongDash}{psline}{linewidth=.0015,linestyle=dashed,dash=.02 .01}
\newpsobject{PST@Diamond}{psdots}{linewidth=.001,linestyle=solid,dotstyle=square,dotangle=45}
\newpsobject{PST@Filldiamond}{psdots}{linewidth=.001,linestyle=solid,dotstyle=square*,dotangle=45}
\newpsobject{PST@Cross}{psdots}{linewidth=.001,linestyle=solid,dotstyle=+,dotangle=45}
\newpsobject{PST@Plus}{psdots}{linewidth=.001,linestyle=solid,dotstyle=+}
\newpsobject{PST@Square}{psdots}{linewidth=.001,linestyle=solid,dotstyle=square}
\newpsobject{PST@Circle}{psdots}{linewidth=.001,linestyle=solid,dotstyle=o}
\newpsobject{PST@Triangle}{psdots}{linewidth=.001,linestyle=solid,dotstyle=triangle}
\newpsobject{PST@Pentagon}{psdots}{linewidth=.001,linestyle=solid,dotstyle=pentagon}
\newpsobject{PST@Fillsquare}{psdots}{linewidth=.001,linestyle=solid,dotstyle=square*}
\newpsobject{PST@Fillcircle}{psdots}{linewidth=.001,linestyle=solid,dotstyle=*}
\newpsobject{PST@Filltriangle}{psdots}{linewidth=.001,linestyle=solid,dotstyle=triangle*}
\newpsobject{PST@Fillpentagon}{psdots}{linewidth=.001,linestyle=solid,dotstyle=pentagon*}
\newpsobject{PST@Arrow}{psline}{linewidth=.001,linestyle=solid}
\catcode`@=12

\fi
\psset{unit=5.0in,xunit=5.0in,yunit=3.0in}
\pspicture(0.000000,0.000000)(1.000000,1.000000)
\ifx\nofigs\undefined
\catcode`@=11

\PST@Border(0.1170,0.0840)
(0.1320,0.0840)

\PST@Border(0.9470,0.0840)
(0.9320,0.0840)

\rput[r](0.1010,0.0840){-50}
\PST@Border(0.1170,0.1644)
(0.1320,0.1644)

\PST@Border(0.9470,0.1644)
(0.9320,0.1644)

\rput[r](0.1010,0.1644){ 0}
\PST@Border(0.1170,0.2447)
(0.1320,0.2447)

\PST@Border(0.9470,0.2447)
(0.9320,0.2447)

\rput[r](0.1010,0.2447){ 50}
\PST@Border(0.1170,0.3251)
(0.1320,0.3251)

\PST@Border(0.9470,0.3251)
(0.9320,0.3251)

\rput[r](0.1010,0.3251){ 100}
\PST@Border(0.1170,0.4055)
(0.1320,0.4055)

\PST@Border(0.9470,0.4055)
(0.9320,0.4055)

\rput[r](0.1010,0.4055){ 150}
\PST@Border(0.1170,0.4858)
(0.1320,0.4858)

\PST@Border(0.9470,0.4858)
(0.9320,0.4858)

\rput[r](0.1010,0.4858){ 200}
\PST@Border(0.1170,0.5662)
(0.1320,0.5662)

\PST@Border(0.9470,0.5662)
(0.9320,0.5662)

\rput[r](0.1010,0.5662){ 250}
\PST@Border(0.1170,0.6465)
(0.1320,0.6465)

\PST@Border(0.9470,0.6465)
(0.9320,0.6465)

\rput[r](0.1010,0.6465){ 300}
\PST@Border(0.1170,0.7269)
(0.1320,0.7269)

\PST@Border(0.9470,0.7269)
(0.9320,0.7269)

\rput[r](0.1010,0.7269){ 350}
\PST@Border(0.1170,0.8073)
(0.1320,0.8073)

\PST@Border(0.9470,0.8073)
(0.9320,0.8073)

\rput[r](0.1010,0.8073){ 400}
\PST@Border(0.1170,0.8876)
(0.1320,0.8876)

\PST@Border(0.9470,0.8876)
(0.9320,0.8876)

\rput[r](0.1010,0.8876){ 450}
\PST@Border(0.1170,0.9680)
(0.1320,0.9680)

\PST@Border(0.9470,0.9680)
(0.9320,0.9680)

\rput[r](0.1010,0.9680){ 500}
\PST@Border(0.1170,0.0840)
(0.1170,0.1040)

\PST@Border(0.1170,0.9680)
(0.1170,0.9480)

\rput(0.1170,0.0420){ 0}
\PST@Border(0.2553,0.0840)
(0.2553,0.1040)

\PST@Border(0.2553,0.9680)
(0.2553,0.9480)

\rput(0.2553,0.0420){ 1}
\PST@Border(0.3937,0.0840)
(0.3937,0.1040)

\PST@Border(0.3937,0.9680)
(0.3937,0.9480)

\rput(0.3937,0.0420){ 2}
\PST@Border(0.5320,0.0840)
(0.5320,0.1040)

\PST@Border(0.5320,0.9680)
(0.5320,0.9480)

\rput(0.5320,0.0420){ 3}
\PST@Border(0.6703,0.0840)
(0.6703,0.1040)

\PST@Border(0.6703,0.9680)
(0.6703,0.9480)

\rput(0.6703,0.0420){ 4}
\PST@Border(0.8087,0.0840)
(0.8087,0.1040)

\PST@Border(0.8087,0.9680)
(0.8087,0.9480)

\rput(0.8087,0.0420){ 5}
\PST@Border(0.9470,0.0840)
(0.9470,0.1040)

\PST@Border(0.9470,0.9680)
(0.9470,0.9480)

\rput(0.9470,0.0420){ 6}
\PST@Border(0.1170,0.9680)
(0.1170,0.0840)
(0.9470,0.0840)
(0.9470,0.9680)
(0.1170,0.9680)

\rput(-.12,0){
\rput[r](0.8200,0.9270){${\cal C}$}
\PST@Solid(0.8360,0.9270)
(0.9150,0.9270)
}

\PST@Solid(0.1170,0.3439)
(0.1170,0.3439)
(0.1308,0.4047)
(0.1447,0.4381)
(0.1585,0.4702)
(0.1723,0.4405)
(0.1862,0.4733)
(0.2000,0.4733)
(0.2138,0.4733)
(0.2277,0.5067)
(0.2415,0.5067)
(0.2553,0.5130)
(0.2692,0.5130)
(0.2830,0.5618)
(0.2968,0.5114)
(0.3107,0.5114)
(0.3245,0.5114)
(0.3383,0.5978)
(0.3522,0.5978)
(0.3660,0.5978)
(0.3798,0.5978)
(0.3937,0.6525)
(0.4075,0.6525)
(0.4213,0.6774)
(0.4352,0.7282)
(0.4490,0.7626)
(0.4628,0.7626)
(0.4767,0.7626)
(0.4905,0.7626)
(0.5043,0.7626)
(0.5182,0.8034)
(0.5320,0.7494)
(0.5458,0.7494)
(0.5597,0.7169)
(0.5735,0.7169)
(0.5873,0.6581)
(0.6012,0.6154)
(0.6150,0.7083)
(0.6288,0.7083)
(0.6427,0.7727)
(0.6565,0.7083)
(0.6703,0.7083)
(0.6842,0.7624)
(0.6980,0.6761)
(0.7118,0.6361)
(0.7257,0.6361)
(0.7395,0.6882)
(0.7533,0.7427)
(0.7672,0.7427)
(0.7810,0.7864)
(0.7948,0.8746)
(0.8087,0.8807)
(0.8225,0.9453)
(0.8363,0.8524)
(0.8502,0.8949)
(0.8640,0.8949)

\rput(-.12,0){
\rput[r](0.8200,0.8850){$100\Delta$}
\PST@Dashed(0.8360,0.8850)
(0.9150,0.8850)
}
\rput(0,0.5891){bits}

\PST@Dashed(0.1170,0.1483)
(0.1170,0.1483)
(0.1308,0.3090)
(0.1447,0.4858)
(0.1585,0.4858)
(0.1723,0.6465)
(0.1862,0.6465)
(0.2000,0.6465)
(0.2138,0.6465)
(0.2277,0.6465)
(0.2415,0.6465)
(0.2553,0.8073)
(0.2692,0.8073)
(0.2830,0.8073)
(0.2968,0.6465)
(0.3107,0.6465)
(0.3245,0.6465)
(0.3383,0.1644)
(0.3522,0.1644)
(0.3660,0.1644)
(0.3798,0.1644)
(0.3937,0.1644)
(0.4075,0.1644)
(0.4213,0.3251)
(0.4352,0.3251)
(0.4490,0.4858)
(0.4628,0.4858)
(0.4767,0.4858)
(0.4905,0.4858)
(0.5043,0.4858)
(0.5182,0.4858)
(0.5320,0.3251)
(0.5458,0.3251)
(0.5597,0.3251)
(0.5735,0.3251)
(0.5873,0.3251)
(0.6012,0.3251)
(0.6150,0.4858)
(0.6288,0.4858)
(0.6427,0.4858)
(0.6565,0.4858)
(0.6703,0.4858)
(0.6842,0.4858)
(0.6980,0.4858)
(0.7118,0.4858)
(0.7257,0.4858)
(0.7395,0.3251)
(0.7533,0.4858)
(0.7672,0.4858)
(0.7810,0.4858)
(0.7948,0.3251)
(0.8087,0.3251)
(0.8225,0.3251)
(0.8363,0.3251)
(0.8502,0.3251)
(0.8640,0.3251)

\PST@Border(0.1170,0.9680)
(0.1170,0.0840)
(0.9470,0.0840)
(0.9470,0.9680)
(0.1170,0.9680)

\catcode`@=12
\fi
\endpspicture

%% file: alife12.bbl
\begin{thebibliography}{}

\bibitem[Bedau et~al., 1998]{Bedau-etal98}
Bedau, M.~A., Snyder, E., and Packard, N.~H. (1998).
\newblock A classification of long-term evolutionary dynamics.
\newblock In Adami, C., Belew, R., Kitano, H., and Taylor, C., editors, {\em
  Artificial Life {VI}}, pages 228--237, Cambridge, Mass. MIT Press.

\bibitem[Caldarelli et~al., 1998]{Caldarelli-etal98}
Caldarelli, G., Higgs, P.~G., and McKane, A.~J. (1998).
\newblock Modelling coevolution in multispecies communities.
\newblock {\em J. Theor, Biol.}, 193:345--358.

\bibitem[Darga et~al., 2008]{Darga-etal08}
Darga, P.~T., Sakallah, K.~A., and Markov, I.~L. (2008).
\newblock Faster symmetry discovery using sparsity of symmetries.
\newblock In {\em Proceedings of the 45st Design Automation Conference},
  Anaheim, California.

\bibitem[Drossel et~al., 2001]{Drossel-etal01}
Drossel, B., Higgs, P.~G., and McKane, A.~J. (2001).
\newblock The influence of predator-prey population dynamics on the long-term
  evolution of food web structure.
\newblock {\em J. Theor. Biol.}, 208:91--107.

\bibitem[G\"ornerup and Crutchfield, 2008]{Gornerup-Crutchfield08}
G\"ornerup, O. and Crutchfield, J.~P. (2008).
\newblock Hierarchical self-organization in the finitary process soup.
\newblock {\em Artificial Life}, 14:245--254.

\bibitem[McKay, 1981]{McKay81}
McKay, B.~D. (1981).
\newblock Practical graph isomorphism.
\newblock {\em Congressus Numerantium}, 30:45--87.

\bibitem[Myrvold and Ruskey, 2001]{Myrvold-Ruskey01}
Myrvold, W. and Ruskey, F. (2001).
\newblock Ranking and unranking permutations in linear time.
\newblock {\em Information Processing Letters}, 79:281--284.

\bibitem[Ray, 1991]{Ray91}
Ray, T. (1991).
\newblock An approach to the synthesis of life.
\newblock In Langton, C.~G., Taylor, C., Farmer, J.~D., and Rasmussen, S.,
  editors, {\em Artificial Life {II}}, page 371. Addison-Wesley, Reading, Mass.

\bibitem[Standish, 1994]{Standish94}
Standish, R.~K. (1994).
\newblock Population models with random embryologies as a paradigm for
  evolution.
\newblock {\em Complexity International}, 2.

\bibitem[Standish, 2000]{Standish98b}
Standish, R.~K. (2000).
\newblock The role of innovation within economics.
\newblock In Barnett, W., Chiarella, C., Keen, S., Marks, R., and Schnabl, H.,
  editors, {\em Commerce, Complexity and Evolution}, volume~11 of {\em
  International Symposia in Economic Theory and Econometrics}, pages 61--79.
  Cambridge UP.

\bibitem[Standish, 2003]{Standish03a}
Standish, R.~K. (2003).
\newblock Open-ended artificial evolution.
\newblock {\em International Journal of Computational Intelligence and
  Applications}, 3:167.
\newblock arXiv:nlin.AO/0210027.

\bibitem[Standish, 2004a]{Standish04a}
Standish, R.~K. (2004a).
\newblock {Ecolab}, {Webworld} and self-organisation.
\newblock In Pollack et~al., editors, {\em Artificial Life IX}, page 358,
  Cambridge, MA. MIT Press.

\bibitem[Standish, 2004b]{Standish04c}
Standish, R.~K. (2004b).
\newblock The influence of parsimony and randomness on complexity growth in
  {Tierra}.
\newblock In Bedau et~al., editors, {\em ALife IX Workshop and Tutorial
  Proceedings}, pages 51--55.
\newblock arXiv:nlin.AO/0604026.

\bibitem[Standish, 2005]{Standish05a}
Standish, R.~K. (2005).
\newblock Complexity of networks.
\newblock In Abbass et~al., editors, {\em Recent Advances in Artificial Life},
  volume~3 of {\em Advances in Natural Computation}, pages 253--263, Singapore.
  World Scientific.
\newblock arXiv:cs.IT/0508075.

\bibitem[Standish, 2010a]{Standish10a}
Standish, R.~K. (2010a).
\newblock Complexity of networks (reprise).
\newblock {\em Artificial Life}.
\newblock submitted. arXiv: 0911.348.

\bibitem[Standish, 2010b]{Standish09a}
Standish, R.~K. (2010b).
\newblock {SuperNOVA}: a novel algorithm for graph automorphism calculations.
\newblock {\em Journal of Algorithms - Algorithms in Cognition, Informatics and
  Logic}.
\newblock submitted, arXix: 0905.3927.

\end{thebibliography}
